\documentclass[
12pt, % The default document font size, options: 10pt, 11pt, 12pt
english, % ngerman for German
singlespacing, % Single line spacing, alternatives: onehalfspacing or doublespacing
twocolumn,
headsepline, % Uncomment to get a line under the header
]{article}

\usepackage{xr}
\usepackage{authblk}
\usepackage{graphicx}% Include figure files
\usepackage{multicol}
\usepackage{amsmath,amssymb}
\usepackage{geometry}
\usepackage{url}
\usepackage{dblfloatfix}
\usepackage{array}
\usepackage{subcaption}
\usepackage{booktabs}
\usepackage{adjustbox,lipsum}  
\usepackage{bibentry}
\usepackage[super,square, comma,sort&compress]{natbib}
\usepackage{float} % Add the float package
\title{Beginner's guide to visual analysis of perovskite and organic solar cell current density-voltage characteristics}
\author[1,2]{Albert These}
\author[3]{L. Jan Anton Koster}
\author[1,4]{Christoph J. Brabec}
\author[1]{Vincent M.~Le Corre\thanks{vincent.le.corre@fau.de}}
\affil[1]{Friedrich-Alexander-Universität Erlangen-Nürnberg (FAU), Materials for Electronics and Energy Technology (i-MEET), Martensstra{\ss}e 7, 91058 Erlangen, Germany}
\affil[2]{Friedrich-Alexander-Universität Erlangen-Nürnberg (FAU), Erlangen Graduate School in Advanced Optical Technologies (SAOT), Paul-Gordan-Str. 6, 91052 Erlangen, Germany}
\affil[3]{Zernike Institute for Advanced Materials, University of Groningen, Nijenborgh 4, 9747 AG Groningen, The Netherlands}
\affil[4]{Helmholtz-Institute Erlangen-Nürnberg (HI ERN), Immerwahrstraße 2, 91058 Erlangen, Germany}

\begin{document}
\twocolumn[
\begin{@twocolumnfalse}
    \maketitle
    \begin{abstract}
        The current density-voltage characteristic (JV) is a critical tool for understanding the behaviour of solar cells. In this article, we present an overview of the key aspects of JV analysis and introduce a user-friendly flowchart that facilitates the swift identification of the most probable limiting process in a solar cell, based mainly on the outcomes of light-intensity-dependent JV measurements. The flowchart was developed through extensive drift-diffusion simulations and a rigorous review of the literature, with a specific focus on perovskite and organic solar cells. Moreover, the flowchart proposes supplementary experiments that can be conducted to obtain a more precise prediction of the primary performance losses. It therefore serves as an optimal starting point to analyse performance losses of solar cells.\\
    \end{abstract}
\end{@twocolumnfalse}
]
% \newpage
\clearpage
\subsection*{Introduction}
\begin{figure}[htp]
    \centering
    \includegraphics[width=\columnwidth]{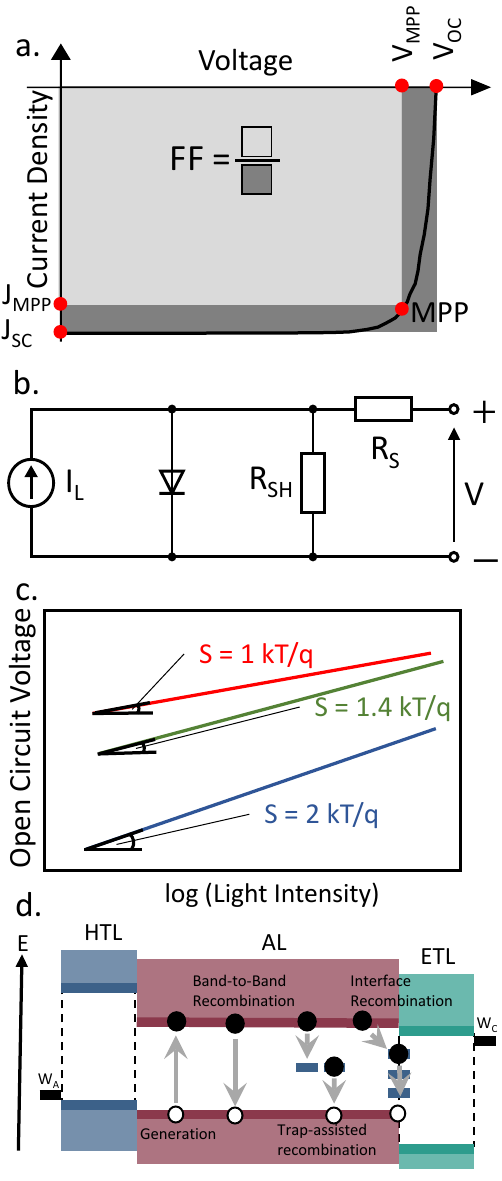}
        \caption{(a) A typical JV curve of a SC showing key parameters. (b) An equivalent circuit for a SC used to model JV curves. (c) Semi-log plot of $V_{OC}$ as a function of light intensity: the slope of the curves can be used to estimate the ideality factor. (d) A schematic band diagram of a SC showing common recombination mechanisms}\label{fig:Intro}
\end{figure}

Current density-voltage characteristic (JVs) are widely acknowledged as the cornerstone in solar cell (SC) research, since they allow for the quantification of a cell’s power conversion efficiency (PCE). However, their significance goes beyond mere efficiency measurements. JVs also provide valuable qualitative insights into the working mechanisms of a SC through careful analysis of their shape and the trends observed in light-intensity-dependent JV measurements.\\
This guide presents a step-by-step approach to analyse SCs and identify performance-limiting factors through the analysis of JV curves. These insights are then utilized to create a flowchart that systematically identifies the primary sources of performance losses in SCs.\\
A typical JV curve is illustrated in Figure \ref{fig:Intro}a. Three key parameters are important to consider when analysing it: the open-circuit voltage ($V_{OC}$), the short-circuit current density ($J_{SC}$) and the fill factor ($FF$). The open-circuit voltage is the voltage at which the net current through the cell is zero and the short-circuit current density is the current density at which the outer cell voltage is zero. The fill factor is defined as:
\begin{equation}\label{eq1:1}
    FF = \frac{J_{MPP}V_{MPP}}{J_{SC}V_{OC}},
\end{equation}
where $J_{MPP}$ and $V_{MPP}$ denote the current density and voltage at the maximum power point (MPP), respectively. Geometrically, it corresponds to the largest rectangle that fits within the JV curve and is defined by the ratio of the areas formed by $J_{SC} * V_{OC}$ and $J_{MPP} * V_{MPP}$. Ultimately, the PCE is defined by:
\begin{equation}\label{eq2:2}
    PCE = \frac{P_{out}}{P_{in}} =\frac{J_{SC}V_{OC}FF}{P_{in}},
\end{equation}
with $P_{in}$ and $P_{out}$ being the incident light power and the output electrical power, respectively.\\
In brief, the $V_{OC}$ is determined by the difference in the quasi-Fermi levels of the electrons and holes in the active layer (AL) of the SC. The $J_{SC}$ is governed by the absorption of the AL and the charge generation rate. The $FF$ is determined by the charge extraction and transport in the AL, transport layers (TLs) and contacts.\\
The equivalent circuit shown in Figure \ref{fig:Intro}b represents a theoretical circuit diagram of a SC and is often used to describe and model the JV. The circuit consists of a current source $I_{L}$ in parallel with a diode. The parallel shunt resistance $R_{SH}$ represents an unwanted current path that diminishes cell performance. $I_{L}$ is connected to a load via a series resistance $R_S$. Such a SC circuit is described by the non-ideal diode equation and can be used to determine the output current  $I$ at voltage $V$:
\begin{equation}\label{eq2:3}
    \begin{split}
        I = & -I_{ph} + I_0 \left[\exp\left(-q\frac{V-R_S I}{nk_bT}\right)-1\right] \\
        & + \frac{V-R_SI}{R_{SH}}
    \end{split}
    \end{equation}
Hereby, $I_{ph}$ is the photocurrent which is generated by the incident light on the SC, $I_0$ is the saturation current related to recombination, $q$ the charge, $k_b$ the Boltzmann constant, $T$ the temperature and $n$ the ideality factor representing the deviation from the ideal diode behaviour.\\
However, such circuit modelling reveals only little insight of the physical processes governing the JV. For example, charge recombination processes are not explicitly included in the circuit model, but only implicitly through the ideality factor $n$. As shown in Figure \ref{fig:Intro}c and described in section III one can empirically determine the ideality factor by doing light intensity dependent JV measurements, but the physical origins are not revealed.\\
Hereto, one has to model the charge carrier recombination explicitly. Recombination events are schematically illustrated in Figure \ref{fig:Intro}d for a hypothetical SC. Charge carriers are generated in the AL and subsequently traverse through drift and diffusion mechanisms. They are then extracted by the TL and contacts or undergo recombination. In this context, recombination can occur radiatively, often involving a bimolecular process that results in photon emission. Alternatively, non-radiative recombination can take place via defect states within the bandgap, referred to as Shockley-Read-Hall (SRH) recombination. This non-radiative process may occur within either the AL or TL, or at the AL/TL interface. The magnitude of all these processes significantly influences the characteristics of the JV in a complex manner and necessitates modelling through drift-diffusion simulations.\\
The evolution and overall shape of such effects on the JV will be the foundation of the flowchart. The flowchart path is determined by answering straightforward questions at each node, leading to the endpoint that identifies the most probable cause of the performance loss. Most of the times, the questions can be answered from analysing the JVs, sometimes additional information about the device is necessary.\\
The decisions at each node are supported by drift-diffusion (DD) simulations as well as references to previous studies for readers wishing for a more in-depth understanding. Occasionally, additional experiments are also suggested to get to a more accurate and definitive prediction of the main losses.\\
The flowchart is designed to be easily understood by a reader. While it will give the right conclusion in most cases it is not excluded that some exotic situations will require a more detailed analysis. The primarily focus is on the analysis of thin-film solar cells like perovskite and organic solar cells. While most of the conclusions and trends discussed here would still hold for more classical photovoltaic technologies such as silicon solar cells, there may be some discrepancies. \\
The flowchart was designed with the support of DD simulations using two representative sets of parameters, one for a typical organic solar cell (OSC) and the other for a perovskite solar cell (PSC). The simulation parameters can be found in supporting information (SI) Tables S1 and S2 with the corresponding fits to real experimental data (Figures S1, S2), taken from ref.\citenum{Lecorre2022} for PSC and \citenum{Hu2020} for OSC.
\subsection*{A few words about drift-diffusion simulations}
Drift-diffusion simulations have been widely used to further understand the device physics of many different solar cell technologies. \cite{Sherkar2017, Koster2005,Neukom2019, Neukom2018} They can be used to model the behaviour  of a device in a wide range of conditions and to quantify the device's performance parameters.\\
Briefly, a DD simulation is a numerical simulation that solves the Poisson and continuity equations as well as the current, drift, and diffusion of charges. Typical input parameters for DD modelling include fundamental semiconductor material properties such as energy levels, mobilities, recombination rate constants, defect/doping densities, etc. \cite{Koster2005,Neukom2019, Neukom2018,Koopmans2022}\\
The details of DD simulations are beyond the scope of this guide and can be found, for example in refs. \citenum{Koster2005,Neukom2018, Koopmans2022}. Nevertheless, we strongly advise readers to explore the literature on DD as, it can offer valuable insights that foster a critical evaluation of experimental results. Numerous papers used DD to investigate the limitations of some classical opto-electrical measurements which we also strongly recommend. \cite{Neukom2019,Neukom2018} One of the advantages DD simulations is the ability the independently investigate the impact of individual parameters, which is rarely possible through experiments alone. We will take advantage of this characteristic of DD simulations when we discuss the impact of the different parameters on the JVs at the different nodes in the flowchart.\\
Here, all the simulations were performed using the open-source drift-diffusion package SIMsalabim.\cite{Koopmans2022,SIMsalabim}, which has an easy to use web interface than can be accessed online.\cite{SIMsalabim_GUI} All of the simulations described in the following can also be reproduced using the Python scripts available on GitHub.\cite{jVBeg}
\subsection*{Flowchart}
\begin{figure*}[ht]
    \centering
    \includegraphics[width=\textwidth]{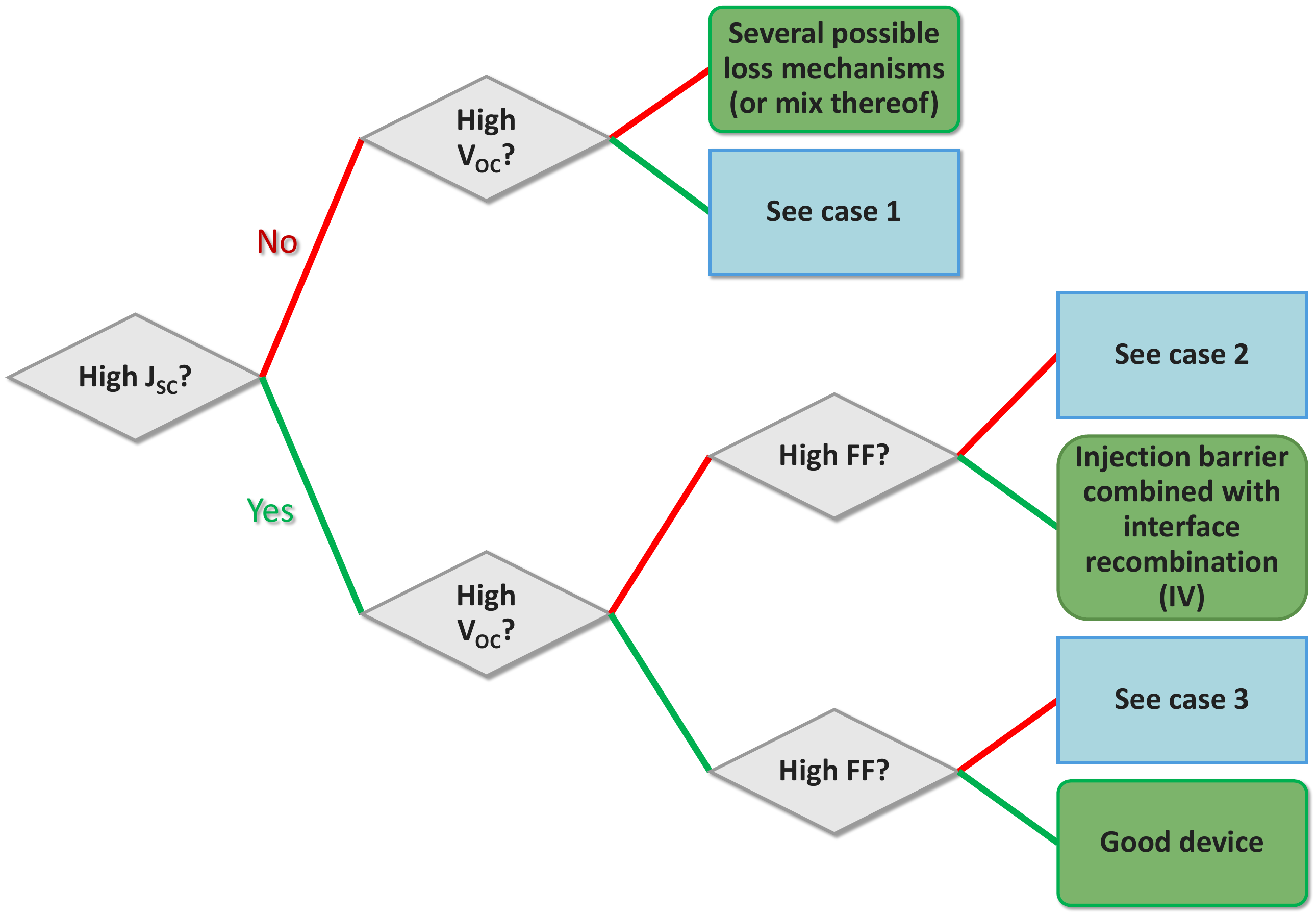}
        \caption{Flowchart to systematically analyse JV characteristics of PSC and OSC.}\label{fig:flowchart}
\end{figure*}

The flowchart is presented in Figure \ref{fig:flowchart}. To ensure clarity, the flowchart will be discussed from left to right and top to bottom. Each endpoint indicating a loss mechanism will be supported by a DD simulation.\\
The grey diamond shaped nodes represent decision nodes where a 'yes-no' question is posed. For clarity, the answer 'yes' is represented by the green line and the 'no' by the red lines. The answers can be directly determined either from the JV, by knowledge about your device, or by performing the suggested experiments. The blue rectangles are process nodes that require an additional action (i.e. supplementary experiments besides JV measurements) that have to be performed before continuing to the next node.\\
To simplify the structure of the flowchart, we have summarized most tasks in several cases. The green rounded rectangles represent the endpoints which should give you the dominant loss. However, it is important to note that reaching an endpoint early in the tree does not necessarily exclude the losses in the subsequent branches. It is therefore useful to explore the tree further down, even if an endpoint has been reached. However, if you do not fall into one of these early endpoints then the corresponding loss mechanism can be excluded. We added the relevant section number to each process and result node in brackets to guide the reader to the corresponding paragraph in the main text and simulations in the supplementary information (SI).\\
In the following, we intentionally refrain from providing any numerical values to define what a good or bad range for the different figures-of-merit is, because such assessments are system dependent. The Shockley-Queisser limit \cite{Shockley1961} provides a good baseline to estimate the loss of the different figure-of-merits. We strongly recommend readers to use online resources with tabulated values\cite{Ruhle2016} or different codes capable of calculating the Shockley-Queisser limit\cite{Chuanggit,Hamadygit} to calculate these losses.\\
However, one must not forget that what is considered a good value also depends a lot on the technology being studied and the reported state-of-the-art performances. For example, the best-in-class OSCs have reached $FF$ values just below 0.82 \cite{Zhang2022}, whereas record values for PSC can each values as high as 0.85.\cite{Shi2023}\\

\begin{figure}[htp]
    \centering
    \includegraphics[width=\columnwidth]{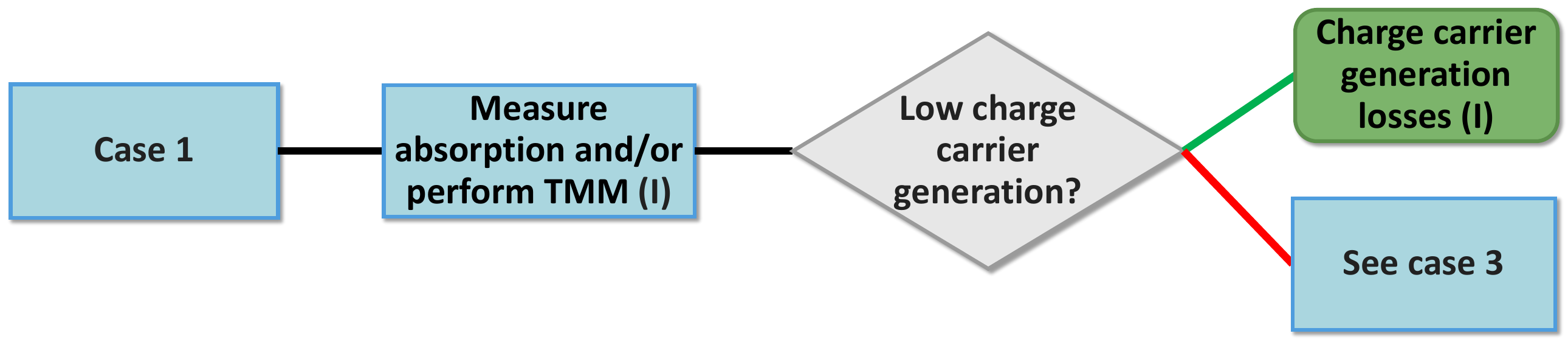}
        \caption{Flowchart illustrating the decision-making process for case 1 with low $J_{SC}$ and high $V_{OC}$}\label{fig:case1}
\end{figure}

Our analysis starts with the $J_{SC}$, as it is the easiest parameter to assess. As can be followed from the flowchart in Figure \ref{fig:flowchart}, a low  $J_{SC}$ and $V_{OC}$ value lead to the worst-case scenario: there are several possible loss mechanisms (or a combination thereof) affecting the cell that are impossible to entangle without additional experiments or information about the device. We discuss the origins of these losses in \textbf{Cases (1-3)} and one can check there, which loss can be excluded for the analysed device.\\
Otherwise, if the $J_{SC}$ is low, but the $V_{OC}$ satisfactory, losses are generally due to the effects describes in \textbf{Case 1} (see Figure \ref{fig:case1}). First, you need to ensure that the AL creates enough photogenerated charge carriers.\\

\textbf{(I) Charge carrier generation losses:}\\
The main parameter that strongly affects the $J_{SC}$ without reducing $V_{OC}$ considerably is the AL absorption and by extension the free-charge carrier generation rate (G). Figures S3-S4 demonstrate the direct correlation between $J_{SC}$, the AL thickness (L) and the average generation rate. In most cases, i.e. in the absence of other potential losses that affect the $J_{SC}$ and are described in the following , it can be expressed as follows:
\begin{equation}\label{eq3:3}
    J_{SC} = qGL,
\end{equation}
with q being the electric charge.\\
In the event of a low $J_{SC}$ the initial consideration is to examine is whether the AL is absorbing enough or not. The absorption of the AL can be assessed by measuring the absorption spectra, although in many cases, a visual inspection is sufficient: a transparent AL suggests that absorption might be limited. A poor absorption is most likely related to either a low absorption coefficient from the AL material itself, or to an AL that is simply too thin.\\
Additionally, a current loss could also be attributed to parasitic absorption from the other layers. This can be checked by performing additional absorption measurements of the other layers or by using transfer matrix modelling to assess the parasitic absorption and the expected $J_{SC}$ from an optical standpoint. We strongly recommend readers to use SIMsalabim \cite{Koopmans2022,SIMsalabim} or the resources provided by the McGehee\cite{Burkhard2010,transfermatrix} and Armin's\cite{Kerremans2020,NKFinder} groups if they wish to perform such simulations. Note that $G$ and hence $J_{SC}$ can be calculated from transfer matrix modelling and using equation~\ref{eq3:3} and, if properly corrected for the internal quantum efficiency, it can be compared to the experimental current to quantify the current loss due to non-optical processes which will be discussed later.\\

\begin{figure*}[htp]
    \centering
    \includegraphics[width=\textwidth]{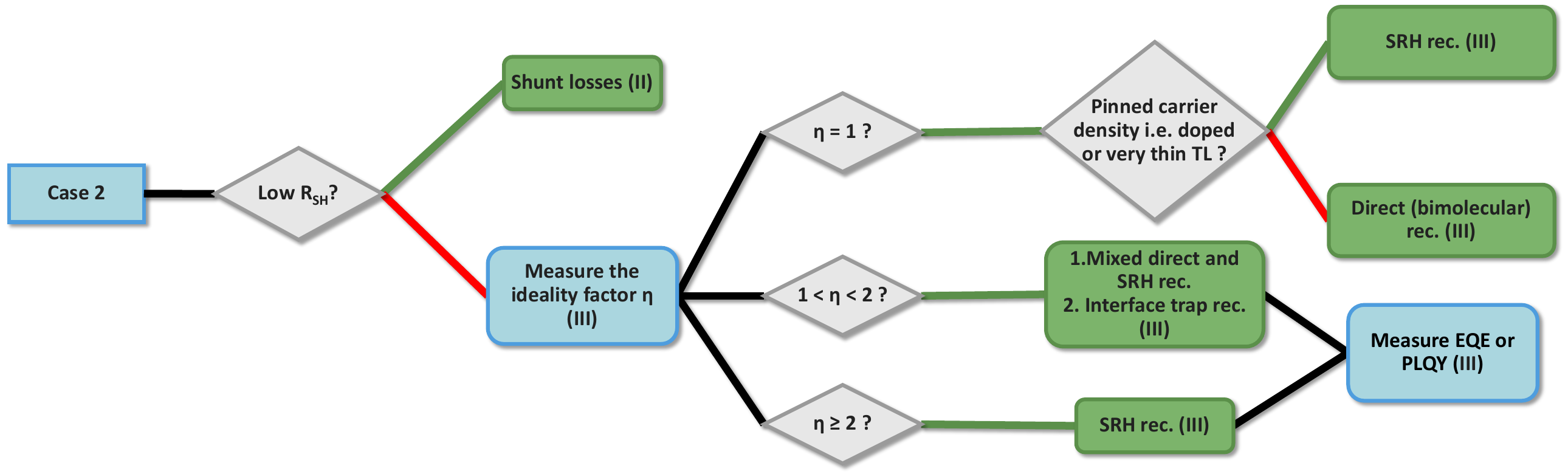}
        \caption{Flowchart illustrating the decision-making process for case 2 with high $J_{SC}$ and low or high $V_{OC}$ and $FF$.}\label{fig:case2}
\end{figure*}

\begin{figure*}[ht]
    \centering
    \includegraphics[width=\textwidth]{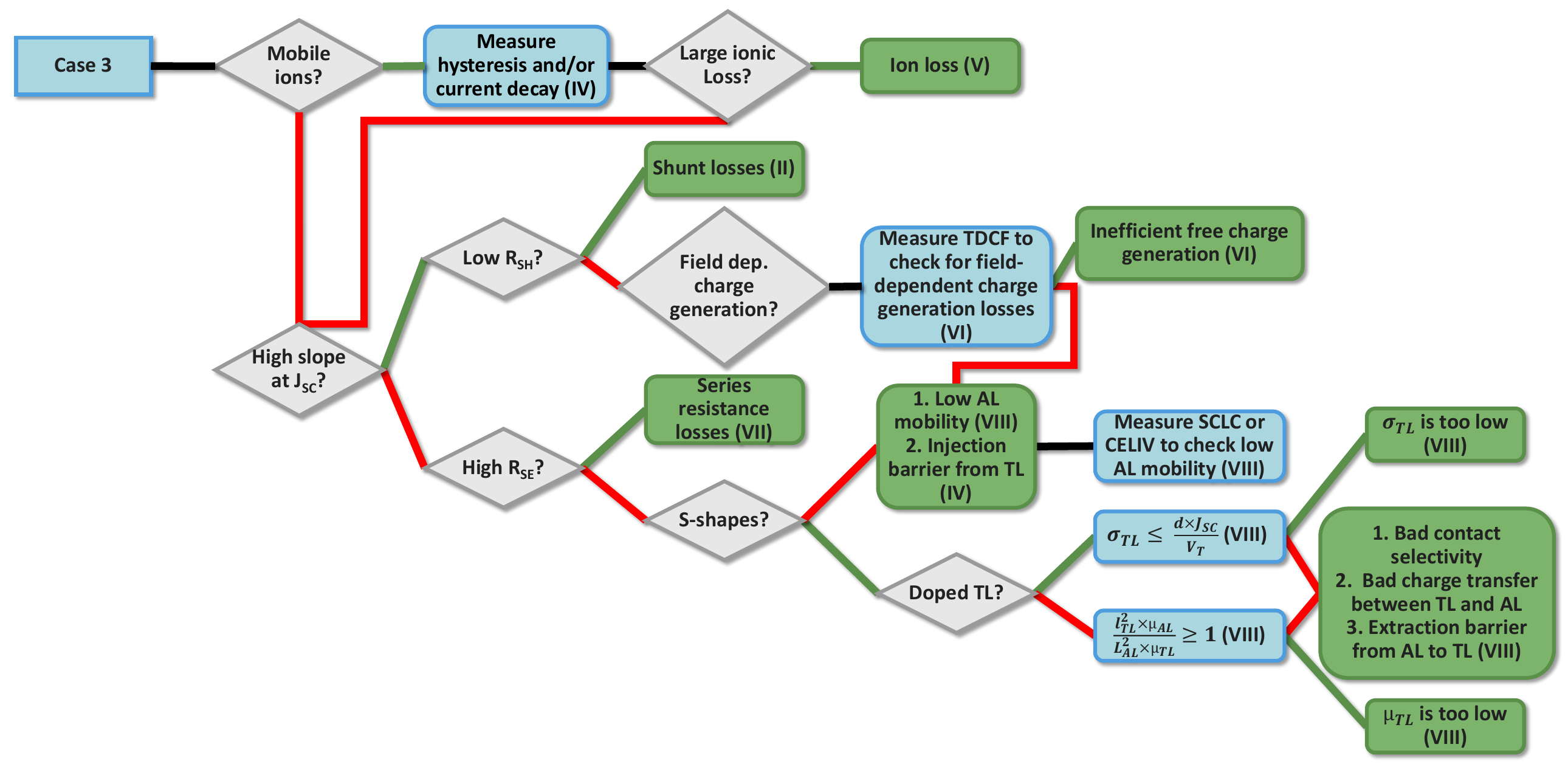}
        \caption{Flowchart illustrating the decision-making process for case 3 with high $J_{SC}$ and $V_{OC}$ and low $FF$.}\label{fig:case3}
\end{figure*}

If low absorption in the AL can be excluded as a reason for a low $J_{SC}$, we fall into \textbf{Case 3} (see Figure \ref{fig:case3}) which is addressed later. However, it should be noted that the losses described in \textbf{Case 3} need to be significant to impact the $J_{SC}$ significantly.\\
We now move on to the bottom half of the tree, where the scenarios having a good $J_{SC}$ are considered. If both the  $V_{OC}$ and $FF$ are low then we fall into \textbf{Case 2} (see Figure \ref{fig:case2}).\\

\textbf{(II) Shunt resistance losses:}\\
The first and easiest loss to check when having a good $J_{SC}$ and bad $FF$ and $V_{OC}$ is the shunt losses. To visually assess if shunts are a dominating factor, one needs to look at the slope at $J_{SC}$. If it is large (under 1 sun illumination) then shunt resistance ($R_{SH}$) is a likely loss. Figures S5-S6 demonstrate that examining the dark-JV at low voltages for high current densities is a good way to judge if $R_{SH}$ is low. Plotting the light intensity dependent $V_{OC}$ and $FF$ is also a useful, and often forgotten, way to check for $R_{SH}$ losses. Typically, if both values significantly decrease at low light intensity, it indicates that shunt losses are significant.
They become increasingly limiting for the performance of the cell if this plummeting regime approaches the 1 sun light intensity.
One can also compare the $J_{SC}$ in dark and under illumination.
Hereby, the empirical formulas shown in equations \ref{shunt_ff} and \ref{shunt_voc} derived in the supporting information can be used to estimate the influence of shunt resistance on the $V_{OC}$ and $FF$ respectively (see Figures S31 and S32).\\
The $FF$ is not detrimentally affected by a low $R_{SH}$, if the following condition is met:
\begin{equation}\label{shunt_ff}
    \frac{J_{SC}}{J_{dark}(-1V)} \gtrsim 100
\end{equation}

Here, $J_{dark}(-1V)$ is the dark current density at $-1$ V (beware of the low breakdown reverse bias of PSC).\\
The $V_{OC}$ is less sensitive to a leakage current and is not detrimentally affected if the following condition applies:

\begin{equation}\label{shunt_voc}
    \frac{J_{SC}}{J_{dark}(-1V)} \gtrsim 5
\end{equation}

The shunt resistance can be quantified with different methods: (a) by taking the slope at $J_{SC}$ with $R_{SH} = |\frac{1}{slope_{J_{SC}}}|$ (hereby, $\frac{J_{SC}}{J_{dark}(-1V)}$ can also be replaced by $J_{SC}*slope_{J_{SC}}$ in eq. \ref{shunt_ff} and \ref{shunt_voc}), (b) by calculating the differential resistance\cite{Brus2016}, or (c) by fitting of the dark-JV (or dark and light-JVs) with the non-ideal diode equation .\cite{Jain2004,Suckow2014,nanohubdoublediode}
The tools by Suckow et al.\cite{Suckow2014,nanohubdoublediode} or by Holmgren et al.\cite{Holmgren2018,pvlibGit} provide a good starting point for fitting the non-ideal diode equation. Note that method (a) can lead to unreliable results since the slope at $J_{SC}$ is affected by other mechanisms (e.g. mobile ions), hence we advise readers to preferably use methods (b) and (c).\\
After excluding shunts as a reason for low $FF$ and $V_{OC}$, it is necessary to examine recombination losses.\\

\textbf{(III) Recombination losses:}\\
One of the easiest methods to check for recombination is to measure light-intensity dependent JVs. As shown in Figure \ref{fig:Intro}c, by plotting the $V_{OC}$ against the logarithm of the light-intensity and performing a linear fit, we can estimate the slope $s$ which depends on the ideality factor $\eta$ such that:
\begin{equation}\label{eq4:4}
    s = \eta \frac{k_BT}{q},
\end{equation}
with $k_B$ being the Boltzmann constant and $T$ the temperature.\\
The most common analysis of the ideality factor limits itself to suggesting that if $\eta$ is close to 1 then band-to-band/bimolecular recombination is the dominant recombination process, which is however not always true. For example, the pinning of one of the carrier densities at the AL/TL interface can lead to dominant SRH recombination with an ideality factor close to 1. The charge carrier density can for example be pinned to the doping level of the TL in case of a defect rich interface and highly doped TL. If the TL is very thin, the charge carrier density can also be pinned to the energy level of the electrode, as it induces plenty of charges in the TL \cite{Lecorre2021ML} These effects are shown in Figures S7-S8 and S13-S14. In the absence of Fermi level pinning, the recombination is bimolecular, see Figures S9 and S15. For PSC, this recombination is radiative and directly correlated to the absorption coefficient as per detailed balance limit \cite{Shockley1961}. It thus will not reduce the performance of the solar cell. However, in our DD simulation, the bimolecular recombination factor is not directly connect to absorption leading to the observed trends. On the other hand, OSCs have processes that result in non-radiative second order recombination\cite{Tvingstedt2016} resulting in the trends shown in S15.
An $\eta$ between 1 and 2 is an indication for either SRH recombination or dominant recombination via interface traps (S10-S12 and S16-S18).
Lastly, $\eta$ can also be larger than 2 for PSC, which indicates dominant SRH recombination, see Figure S10. However, as discussed in more detail in refs.\citenum{Tress2018,Caprioglio2020,Lecorre2021ML} other factors can also influence the ideality factor. Readers need to be careful when drawing strong conclusions from the ideality factor alone.\\
Ideally, we recommend performing additional experiments to confirm the conclusion drawn from analysing the ideality factor alone. For example, intensity-dependent and/or ultra-sensitive external quantum efficiency (EQE) measurements can be a good way to assess the presence of traps in the systems.\cite{Zarrabi2020,Zeiske2021} Measuring quasi-fermi level splitting (QFLS) on the AL material, half-cells with the TLs and the AL and complete cells is also a very powerful way to pinpoint where the dominant recombination happens.\cite{Phuong2021a, Stolterfoht2019a} However, depending on the technologies, different methods need to be used to measure the QFLS. For PSCs absolute photoluminescence (PL) and quantum yield measurements are routinely used since the PL mostly depends on the recombination of free carriers.\cite{Stolterfoht2018,Kirchartz2020} For OSCs, alternative methods, such as photoinduced absorption spectroscopy are suitable.\cite{Phuong2021} In fact, the presence of exciton and charge-transfer states in organic materials and blends strongly influence the PL making it an unsuitable method to estimate the QFLS. For more insights into PL and QFLS measurements we recommend reading refs.\citenum{Stolterfoht2018,Kirchartz2020,Phuong2021,Kuckemeier2021,Stolterfoht2019,Stolterfoht2019b,Wolff2019}.\\

\textbf{(IV) Injection barrier and trapping at the transport layer to active layer interface losses:}\\
The next endpoint occurs when one observes a high $J_{SC}$ combined with a low $V_{OC}$ and a relatively high $FF$. This is a typical signature of an injection barrier from the TL to the AL combined with significant SRH recombination occurring  at that interface. Figures S20-S21 a-b demonstrate this phenomena. It can be investigated by comparing the QFLS and the $V_{OC}$. If they do not match, it indicates the existence of a significant injection barrier and interface recombination. This mismatch occurs due to a pinning of the QFLS at the interface which limits the $V_{OC}$. The papers by Phuong\cite{Phuong2021} and Stolterfoht\cite{Stolterfoht2019} describe in detail how to conduct these measurements for OSC and PSC respectively.\\
Even in the absence of dominant traps, the presence of an injection barrier can  hinder efficient charge transfer and extraction, leading to a reduction of $FF$ as shown in S20 and S21 c-d.\cite{Stolterfoht2019a}\\ However, this is only happens if the injected charge carrier mobility is low in the AL. Otherwise, an injection barrier does not affect the SC detrimentally (S20 e-f). The instances of injection barriers in absence of SRH recombination do not reduce the $V_{OC}$ and thus are treated under \textbf{Case 3}.\\
If both the $J_{SC}$ and $V_{OC}$ are high, but the $FF$ is low then we fall  into \textbf{Case 3}.\\
\\\\
\textbf{(V) Ionic losses:}\\
The first loss mechanism to consider in \textbf{Case 3} are ionic losses. As demonstrated in Figure S22 a-b, the presence of a high concentration of both positively and negatively charged mobile ions significantly affects the $FF$ of PSCs, while retaining a high $J_{SC}$ and $V_{OC}$. However, if the TLs are non-blocking and ions move inside the TLs, these effects are strongly mitigated and the mobile ions influence the JV less S22 c-d. Note that these mobile ions may then induce detrimental effects on long-term stability and degrade both the TLs and electrodes which affects the performance of the device over time \cite{Kerner2019,Bitton2023}.\\
If only one polarity type is dominant, mobile ions can also lead to a low $FF$ (Figure S23 a-b) or to a low $J_{SC}$ (Figure S23 c-d). Non-blocking TLs again reduce the influence of both effects (not shown here). Thiesbrummel et al.,\cite{Thiesbrummel2021} demonstrated that the ions decrease the $J_{SC}$ by reducing the charge extraction efficiency caused by the ionic accumulation at the interface between the perovskite and TL and flattening of the bands under stabilized short-circuit condition.  To assess whether ionic losses are significant, one needs to perform either fast hysteresis measurements as described in ref.\citenum{Lecorre2022} or measure the transient current decay as described in ref.\citenum{Thiesbrummel2021}.\\ 
If ionic losses can be excluded then one has to examine the slope at $J_{SC}$. If it is high, we have to ensure that we do not have any shunt losses, as already described in section II.\\

\textbf{(VI) Field-dependent charge carrier generation losses:}\\
 Another common phenomena that can lead to a high slope at $J_{SC}$ for OSC includes field-dependant charge carrier generation.\cite{Kurpiers2018} Small highest occupied molecular orbital to lowest unoccupied molecular orbital energetic offsets can lead to inefficient dissociation of excitons into charge transfer states. The dissociation is then field dependent and can lead to a reduction of the $J_{SC}$ and $FF$, see Figure S24.\cite{Pranav2023, Tokmoldin2023} To properly investigate if this process is causing the slope at $J_{SC}$, one needs to perform more advanced experimental techniques such as time-delayed collection field (TDCF)\cite{Kurpiers2018} and/or light-intensity and voltage dependent EQE.\cite{Wurfel2019}\\ If field-dependant charge carrier generation is not an issue, then the AL mobility is likely the limiting factor, as discussed in section VIII.
\\
\\
\textbf{(VII) Series resistance losses:}\\
Otherwise, if the slope is small, series resistance losses need to be examined. Similarly to the shunt resistance, the series resistance can be estimated with different methods: (a) by taking the slope at $V_{OC}$ with $R_S = |\frac{1}{slope_{OC}}|$, (b) by calculating the differential resistance\cite{Brus2016} or (c) by fitting of the dark-JV (or dark and light-JVs) with the non-ideal diode equation.\cite{Jain2004,Suckow2014,nanohubdoublediode,LeCorreDiode}\\
Assessing the $FF$ vs logarithm of the light intensity is good to determine if $R_S$ strongly affects the performance at 1 sun, see Figures S25-S26. At high light intensities the $FF$ typically decays due to $R_S$, as also shown in Figure S31a. If that decay starts before 1 sun then the device likely suffers from strong series resistance losses.\\
\\
\textbf{(VIII) Extraction and transport losses:}\\
If all resistive issues can be excluded, the next step is to check for S-shapes. A JV-curve is said to have an S-shape, if it has an inflection point, see i.e. Figures S28 and S30. In the absence of S-shapes low electron and hole mobilities in the AL are probably responsible for the low $FF$, as illustrated in Figures S27 and S29. Note that for a given mobility, the AL thickness could also be too large, since the mobility-thickness ratio is the governing parameter here. To check for low AL mobility, one can calculate the effective mobility using the charge extraction by linearly increasing voltage (CELIV) measurement\cite{Neukom2011,Sandberg2019}. Although it does not provide separate values for electron and hole mobilities independently, it still offers valuable information on the charge transport within the device. Another option is to perform space-charge-limited current (SCLC) measurements in a single-carrier device configuration. SCLC measurements are powerful as we can measure the electron and hole mobilities (and trap densities) separately. However, it requires the preparation of additional devices with a different device structure, which would entail further optimisation to assure comparability of the deposited materials. In addition, making reliable SCLC measurements and extracting the correct values from this method is not always simple, we suggest readers to read refs.\citenum{Blakesley2014,Rohr2018Explor,Rohr2019,Duijnstee2020toward,Lecorre2021SCLC} for more information about SCLC.\\
An injection barrier from the TL to the AL can also lead to a $FF$ loss without significant $J_{SC}$ or $V_{OC}$ losses, as already shown in Figure S20-S21a-b. However, this only happens if there is no strong SRH recombination at that interface as describes in section IV.\\
Finally, S-shapes usually appear due to an unbalanced extraction of charge carriers. Such a phenomenon can have several origins such as (a) poor TL mobility or (b) conductivity, (c) poor charge carrier transfer between the AL and the TLs, (d) injection or (e) extraction barriers from the AL to the TLs. One of the current authors previously described how to optimize the TL thickness, mobility and/or conductivity to avoid such losses and define the two figures-of-merit that appear in the flowchart.\cite{Lecorre2019TL}\\

Finally, if the $J_{SC}$, $V_{OC}$ and $FF$ are all high, one has a well performing solar cell.

\section*{Conclusion}
We have presented a list of the most common loss mechanisms in PSC and OSC and showed, using DD simulations, how they affect JVs. We think that the results of this study will be useful for individuals new to the field of thin film SCs. We aim to provide beginners with a framework to visually assess and understand performance losses of solar cells only by looking at JV characteristics. Additionally, a series of references with useful resources and with more in-depth analysis of the different mechanisms is provided here. Readers are encouraged to make use of the many open-source simulation tools available online to deepen their understanding of the different mechanisms and to improve their analysis of experimental data.\\
Finally, it is important to emphasize that the flowchart presented here is only made for a qualitative assessment of the dominant losses and is by no means quantitative. Nevertheless, it is a valuable starting point for the analysis of experimental data. It provides guidance on the experiment to perform next in order to obtain a definitive answer and/or quantify the main loss process. By following the suggested path in the flowchart, researchers can systematically investigate and understand the underlying mechanisms that contribute to performance limitations in their solar cells.\\

\section*{Acknowledgements}
The authors would like to express their gratitude to S. Kahmann for the valuable feedback provided. A.T. gratefully acknowledges funding of the Erlangen Graduate School in Advanced Optical Technologies (SAOT) by the Bavarian State Ministry for Science and Art.\\

\bibliographystyle{vlcref}
\bibliography{guide}
\end{document}

\maketitle
\clearpage

\section*{Solar cell base cases:}

The base case scenario for perovskite solar cells (PSCs) was taken from ref.\citenum{Lecorre2022} and fitted with SIMsalabim.\cite{Koopmans2022,SIMsalabim}
\begin{figure*}[!htb]
    \centering
    \includegraphics[width=\textwidth]{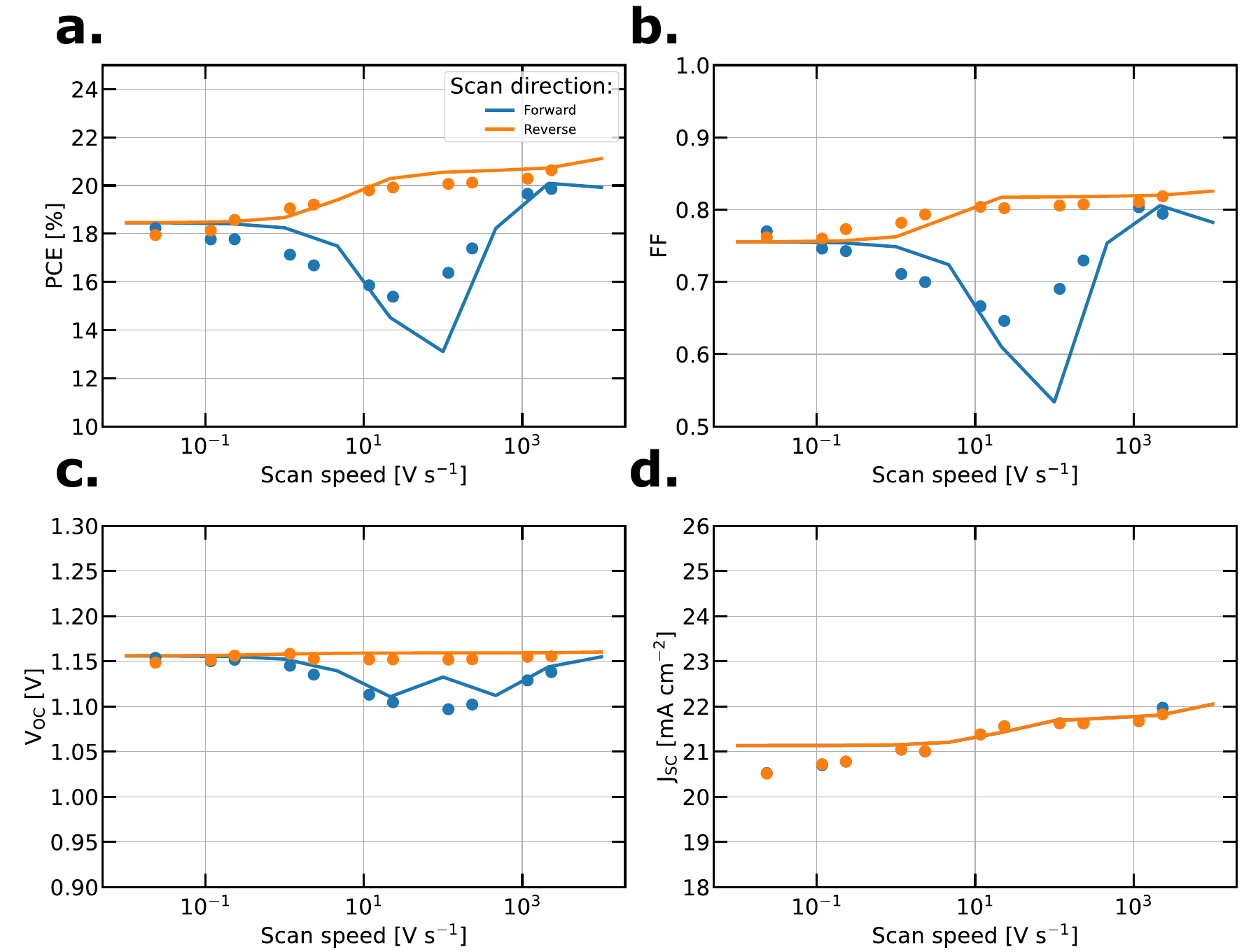}
        \caption{Performances characteristics for the base case scenario for a typical PSC with experimental data (symbols) and the corresponding fitted simulation data (lines). The PCE (a), $FF$ (b), $V_{OC}$ and $J_{SC}$ are plotted as a function of bias scan speed.}\label{fig:Pero_base_case}
\end{figure*}\FloatBarrier

\clearpage

The base case scenario for organic solar cells (OSCs) was taken from ref.\citenum{Hu2020} and fitted with SIMsalabim.\cite{Koopmans2022,SIMsalabim}
\begin{figure*}[!htb]
    \centering
    \includegraphics[width=\textwidth]{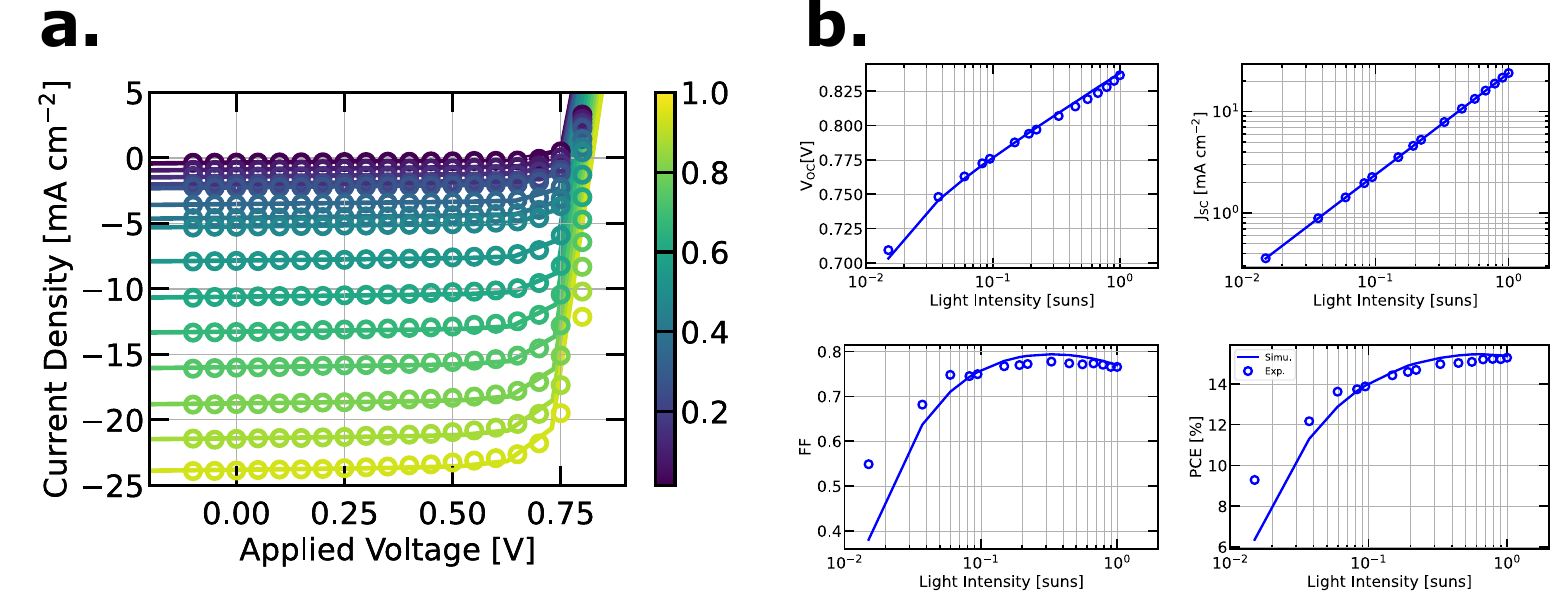}
        \caption{JVs and performances for the base case scenario for a typical OSC with experimental data (symbols) and the corresponding fitted simulation data (lines).  (a) light intensity dependent JV. The colourmap indicates the illumination intensity in fractions of 1 sun. (b) light intensity dependent performances}\label{fig:OSC_base_case}
\end{figure*}\FloatBarrier

\clearpage

\section{Charge carrier generation losses:}
\subsection*{PSC:}
\begin{figure*}[!htb]
    \centering
    \includegraphics[width=\textwidth]{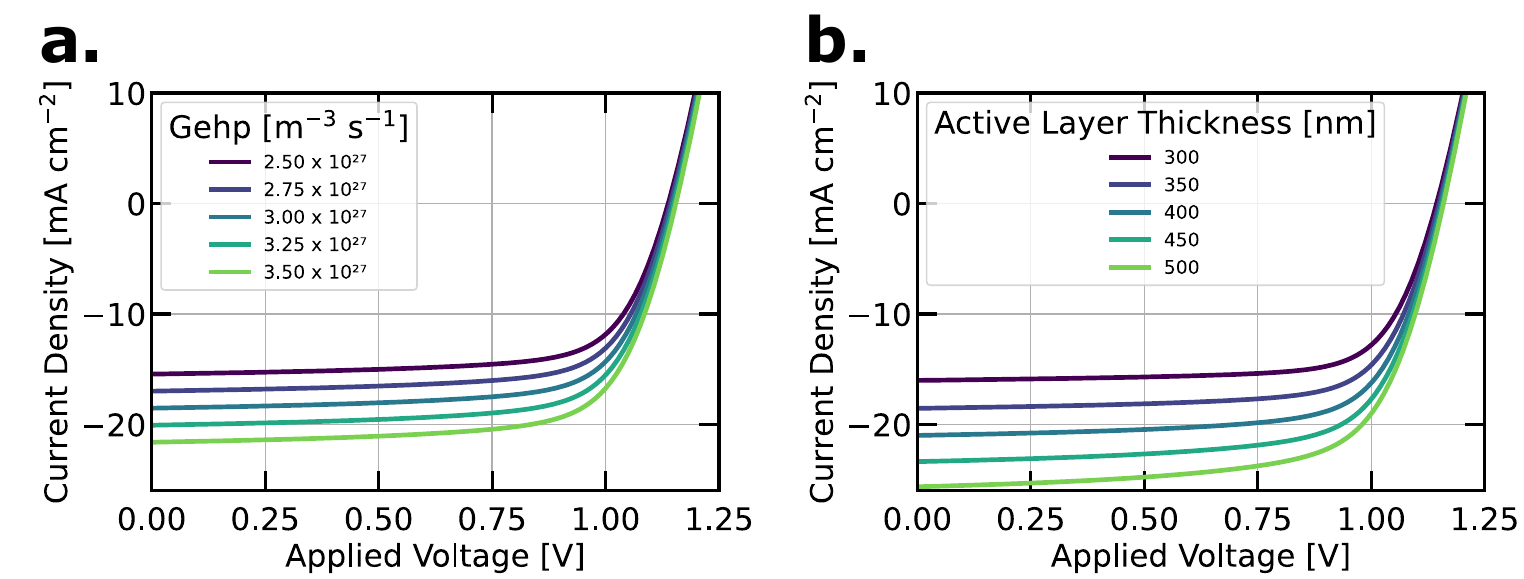}
        \caption{Influence of the (a) electron-hole pair generation rate and (b) active layer thickness on the JVs}\label{fig:abs}
\end{figure*}\FloatBarrier

\subsection*{OSC:}
\begin{figure*}[!htb]
    \centering
    \includegraphics[width=\textwidth]{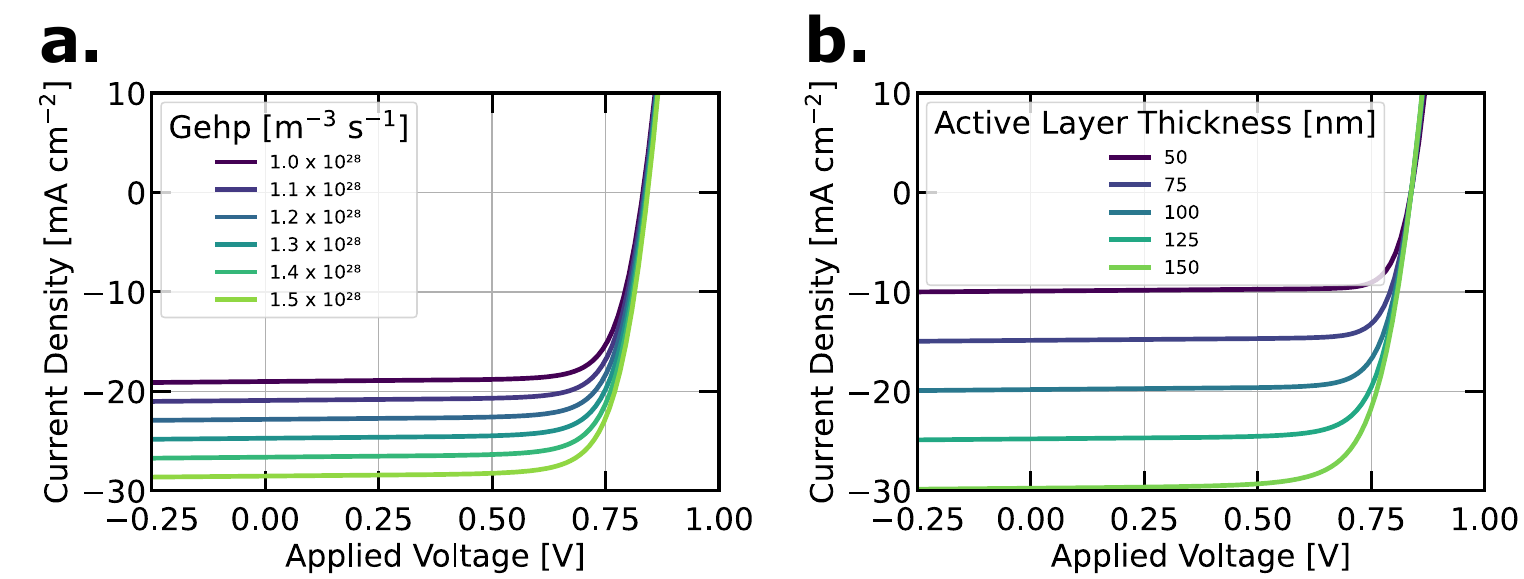}
        \caption{Influence of the (a) electron-hole pair generation rate and (b) active layer thickness on the JVs}\label{fig:OSC_abs}
\end{figure*}\FloatBarrier
\clearpage

\section{Shunt resistance losses:}
\subsection*{PSC:}
\begin{figure*}[!htb]
    \centering
    \includegraphics[width=\textwidth]{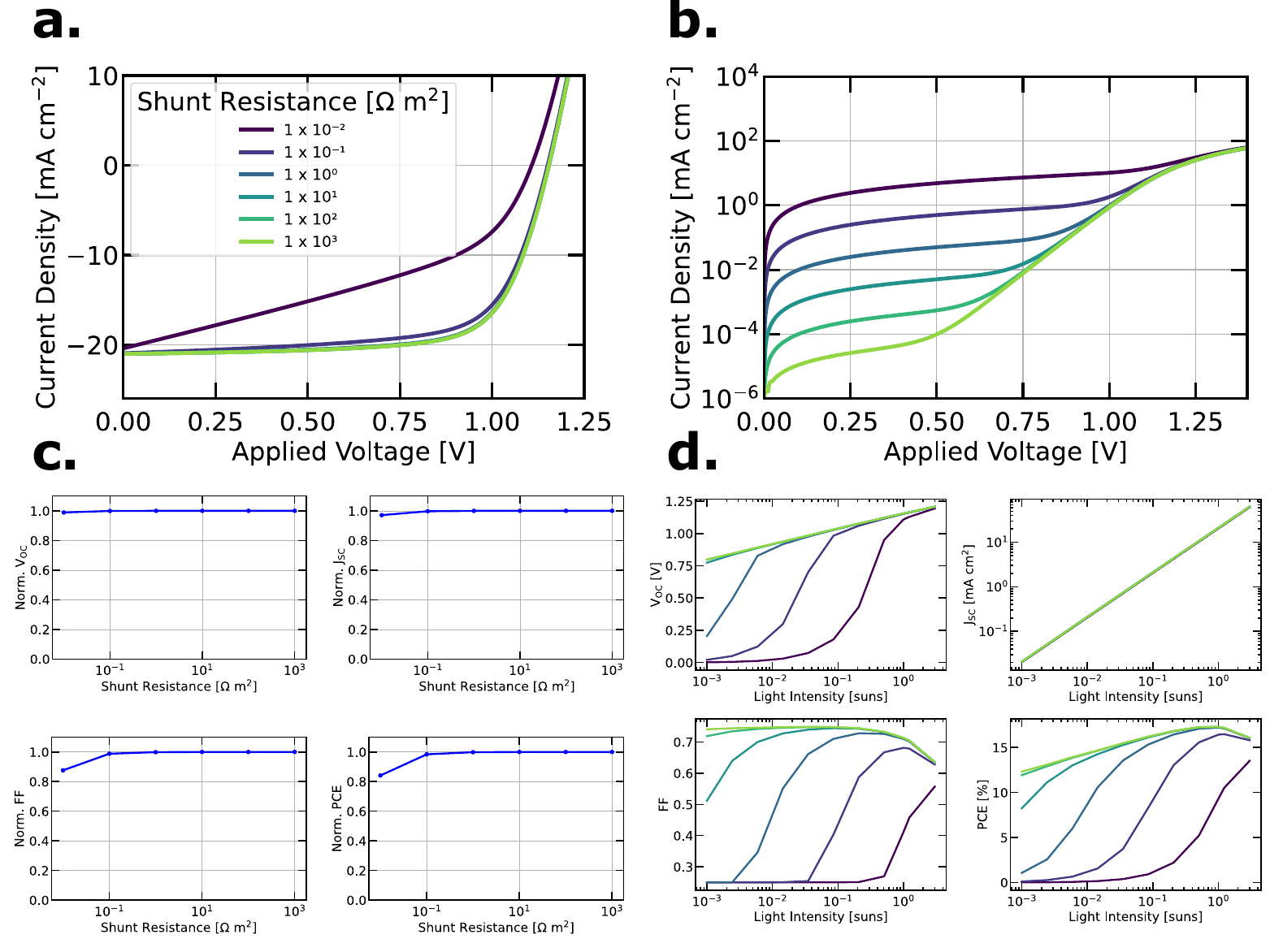}
        \caption{Influence of shunt resistance on the (a) light and (b) dark JVs, (c) 1 sun and (d) light-intensity dependent performances.}\label{fig:shunt}
\end{figure*}\FloatBarrier

\clearpage
\subsection*{OSC:}
\begin{figure*}[!htb]
    \centering
    \includegraphics[width=\textwidth]{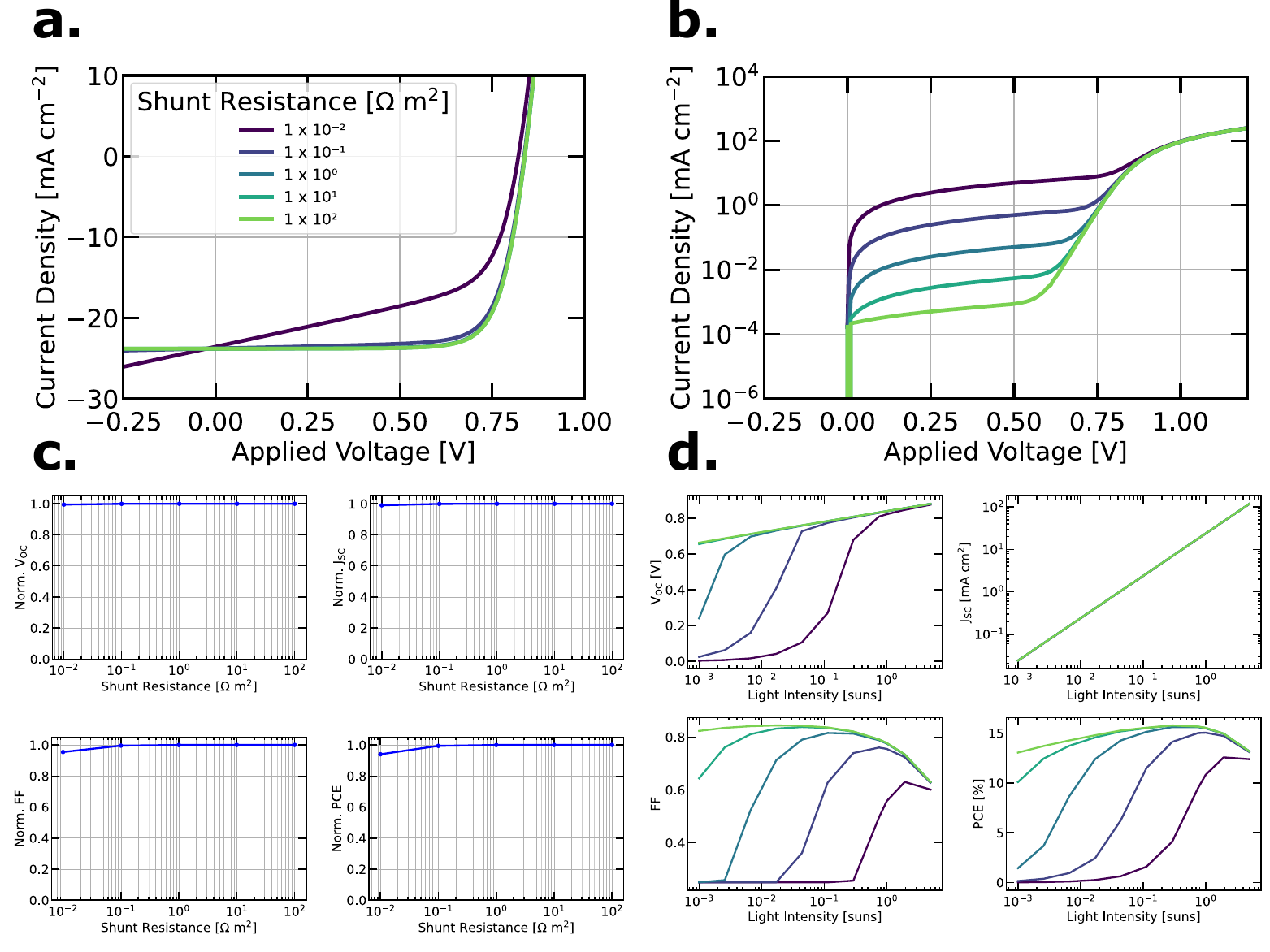}
        \caption{Influence of shunt resistance on the (a) light and (b) dark JVs, (c) 1 sun and (d) light-intensity dependent performances.}\label{fig:OSC_shunt}
\end{figure*}\FloatBarrier

\clearpage
\section{Recombination losses:}
\subsection*{PSC:}
\begin{figure*}[!htb]
    \centering
    \includegraphics[width=\textwidth]{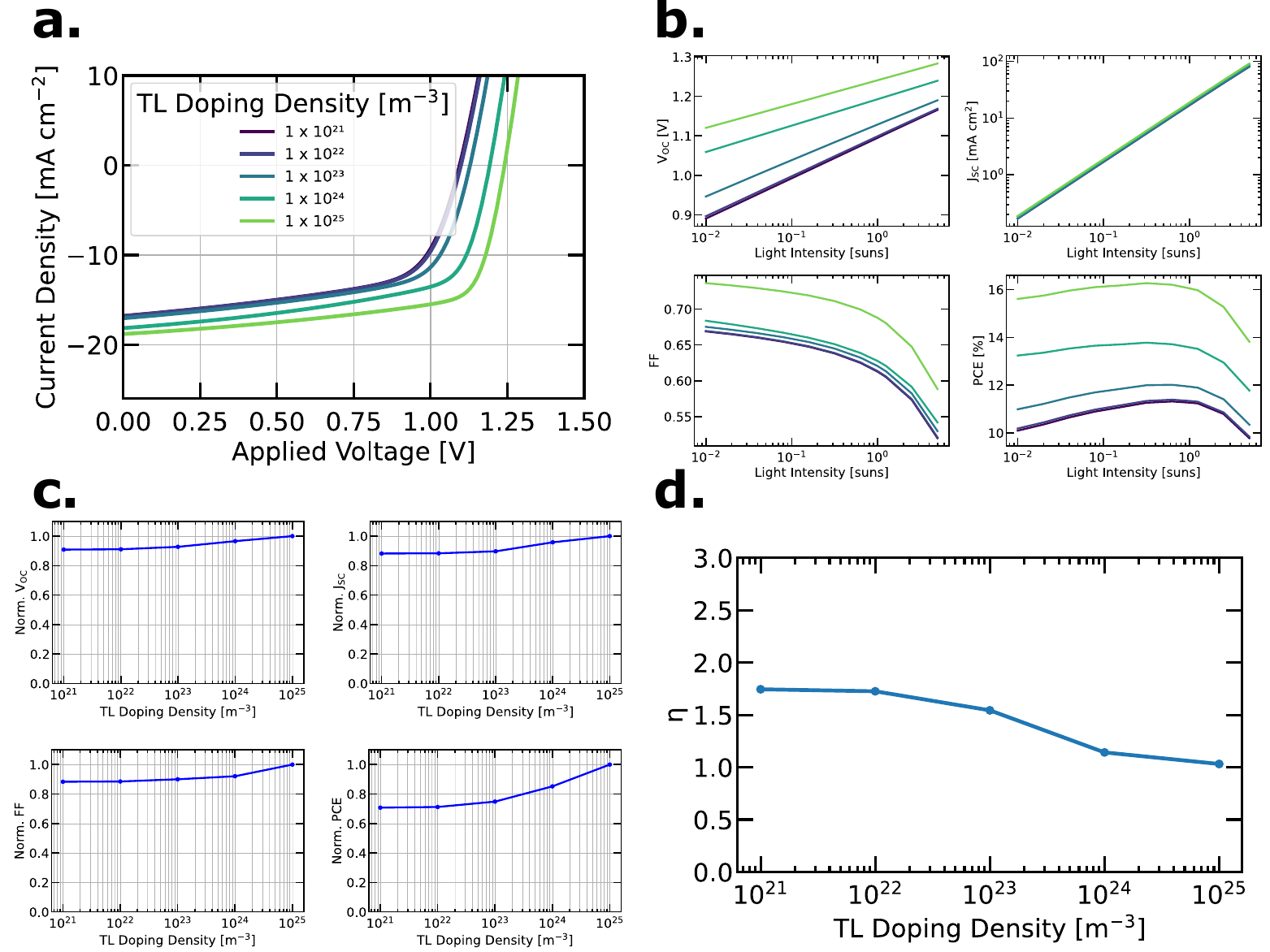}
        \caption{Influence of Fermi level pinning due to doping in the TL on the (a) light JVs, (b) light-intensity dependent and (c) 1 sun performances, (d) ideality factor.}\label{fig:pcd_dop_tl}
\end{figure*}
\begin{figure*}[!htb]
    \centering
    \includegraphics[width=\textwidth]{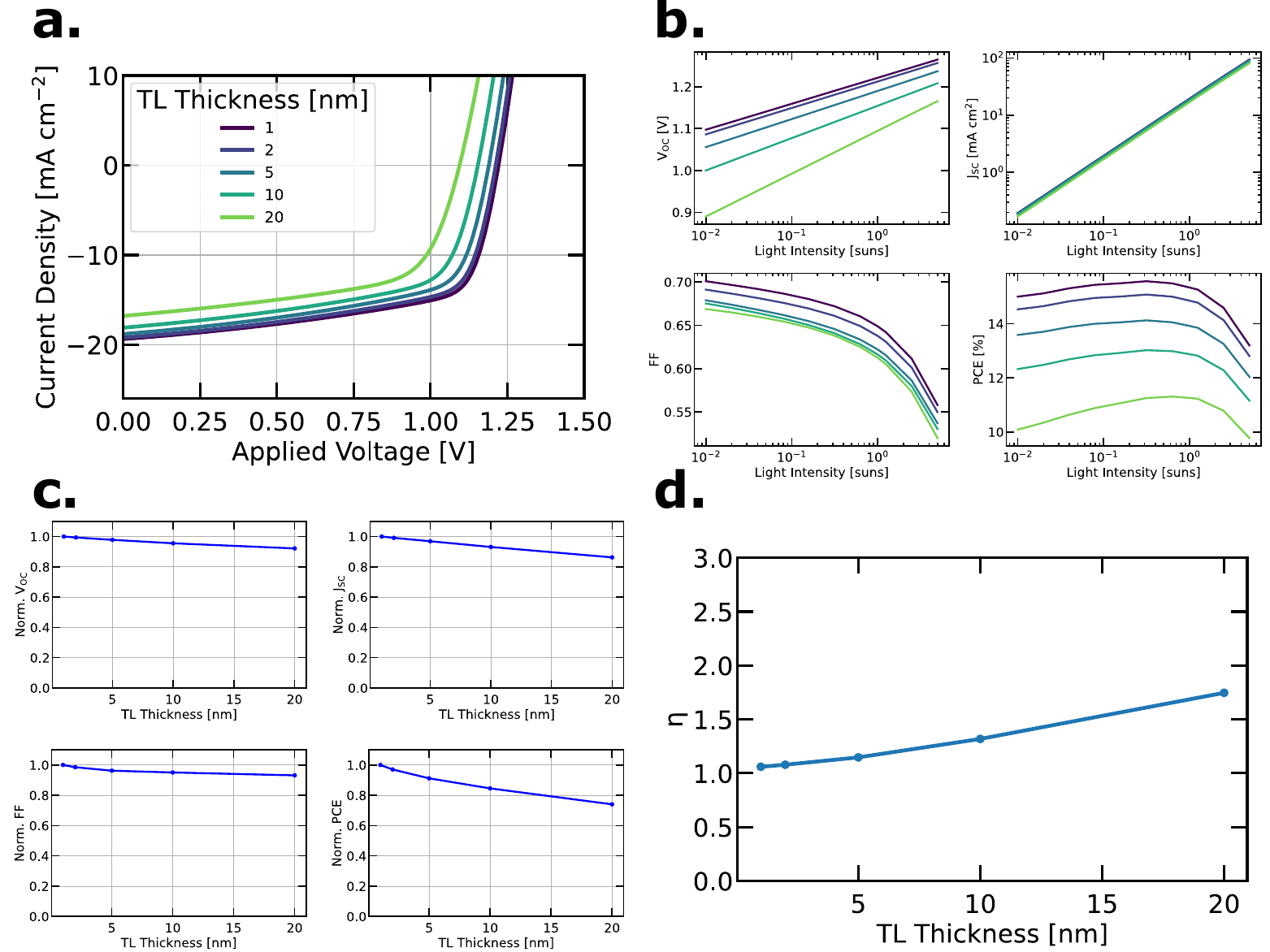}
        \caption{Influence of Fermi level pinning due TL thickness on the (a) light JVs, (b) light-intensity dependent and (c) 1 sun performances, (d) ideality factor.}\label{fig:pcd_l_tl}
\end{figure*}
\begin{figure*}[!htb]
    \centering
    \includegraphics[width=\textwidth]{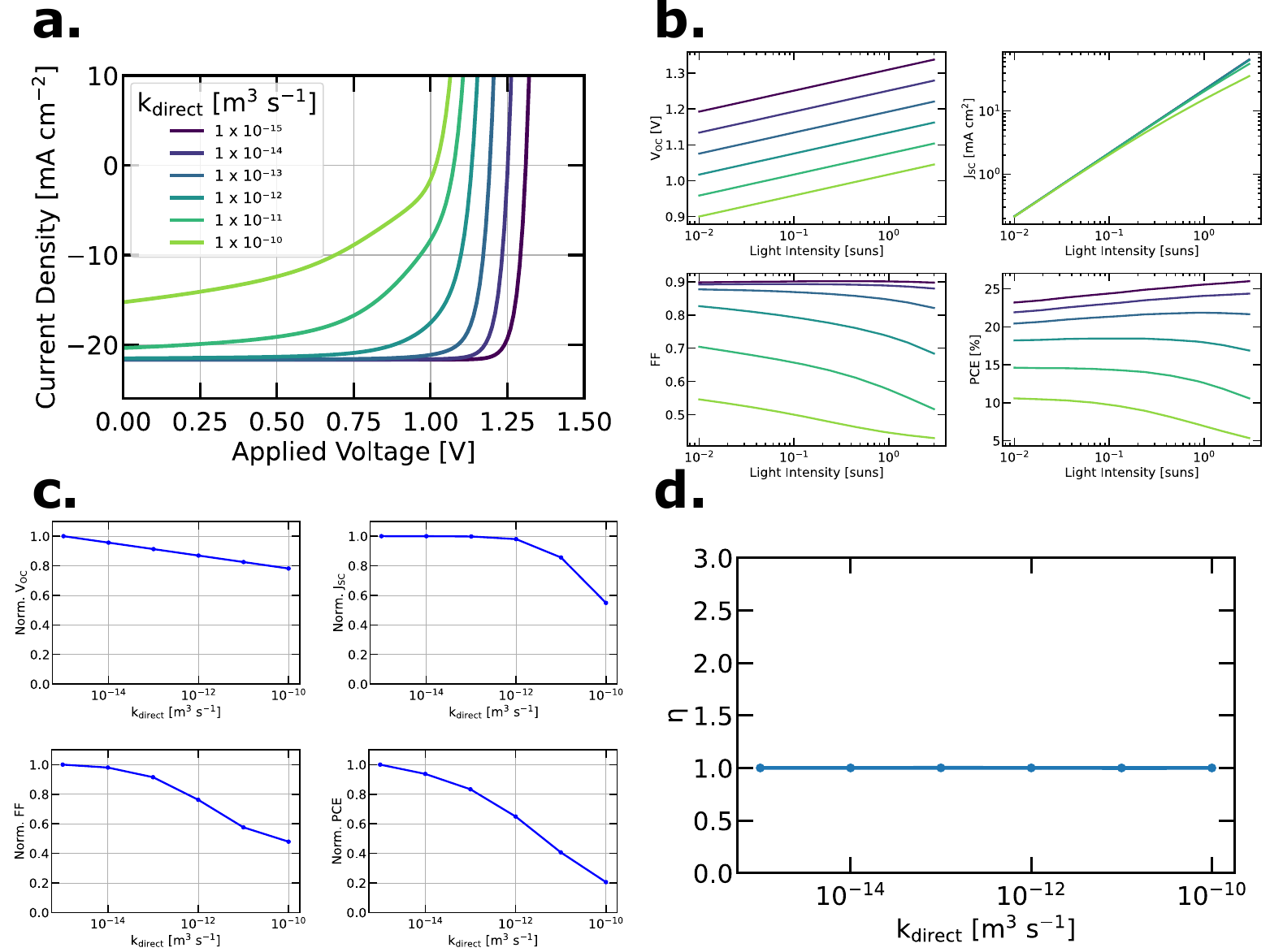}
        \caption{Influence of the band-to-band/bimolecular recombination rate ($k_2$) on the (a) light JVs, (b) light-intensity dependent and (c) 1 sun performances, (d) ideality factor.}\label{fig:k2}
\end{figure*}
\begin{figure*}[!htb]
    \centering
    \includegraphics[width=\textwidth]{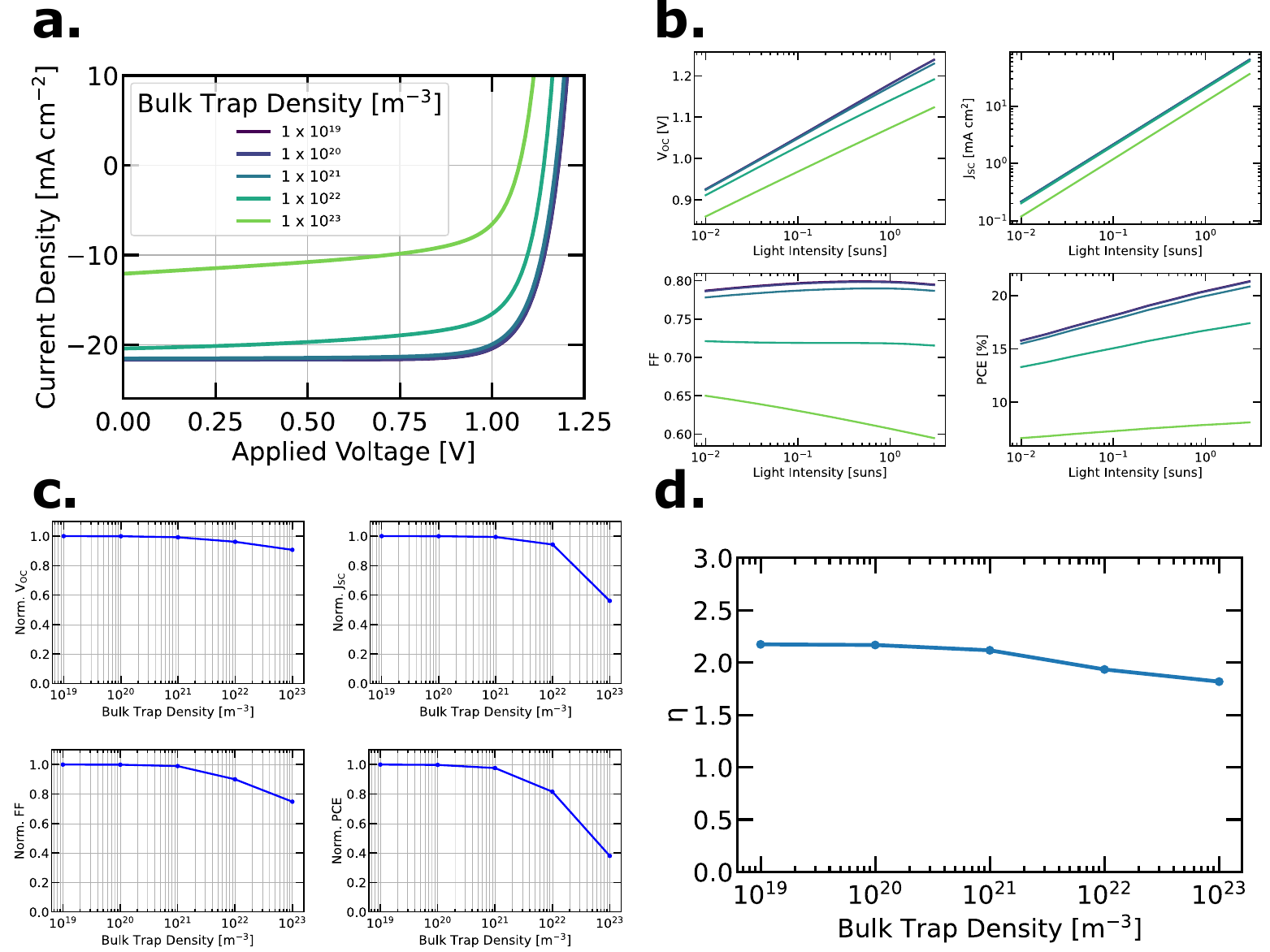}
        \caption{Influence of the bulk trap density on the (a) light JVs, (b) light-intensity dependent and (c) 1 sun performances, (d) ideality factor.}\label{fig:trap}
\end{figure*}
\begin{figure*}[!htb]
    \centering
    \includegraphics[width=\textwidth]{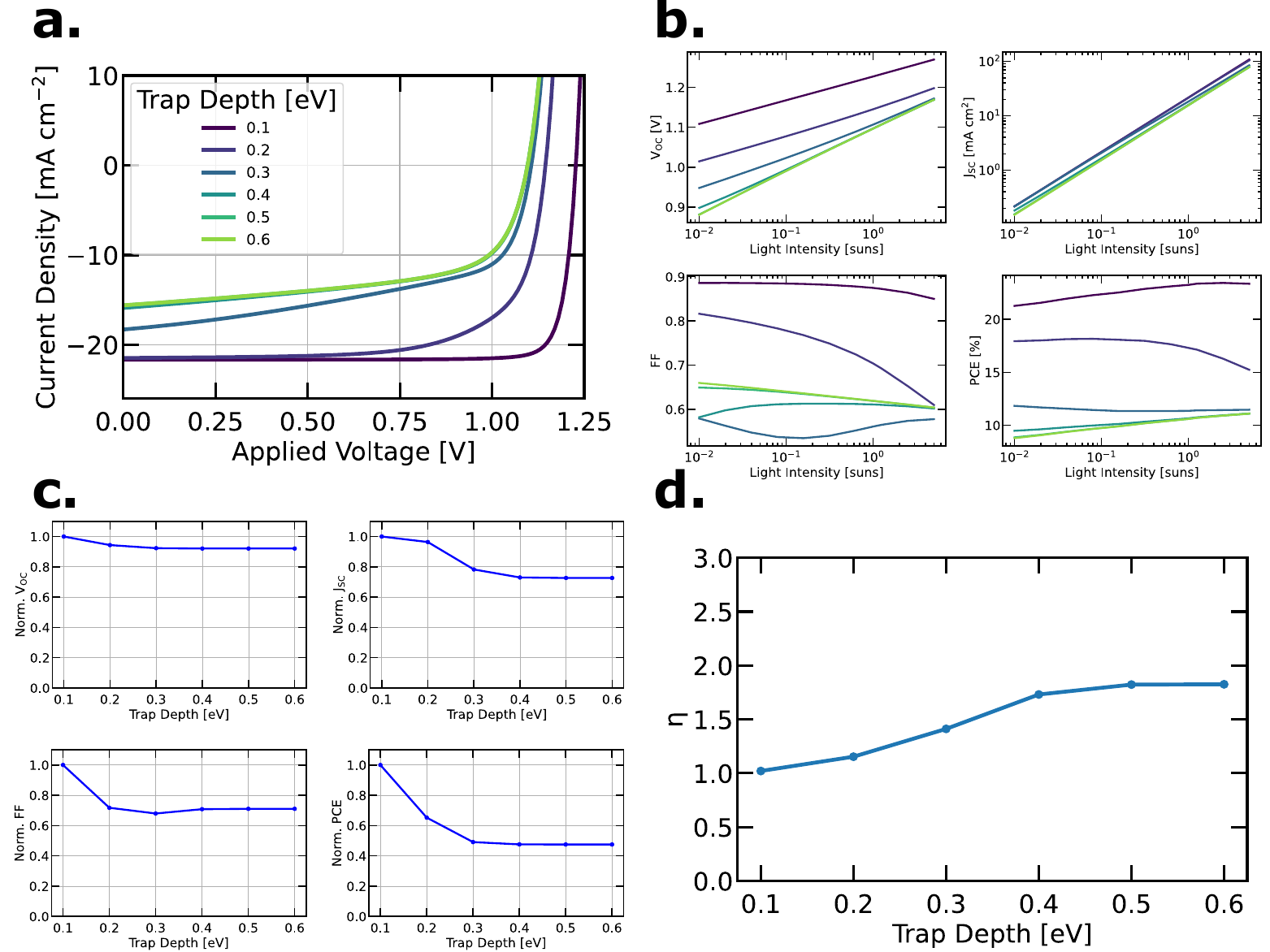}
        \caption{Influence of the trap level depth on the (a) light JVs, (b) light-intensity dependent and (c) 1 sun performances, (d) ideality factor.}\label{fig:trap_depth}
\end{figure*}

\begin{figure*}[!htb]
    \centering
    \includegraphics[width=\textwidth]{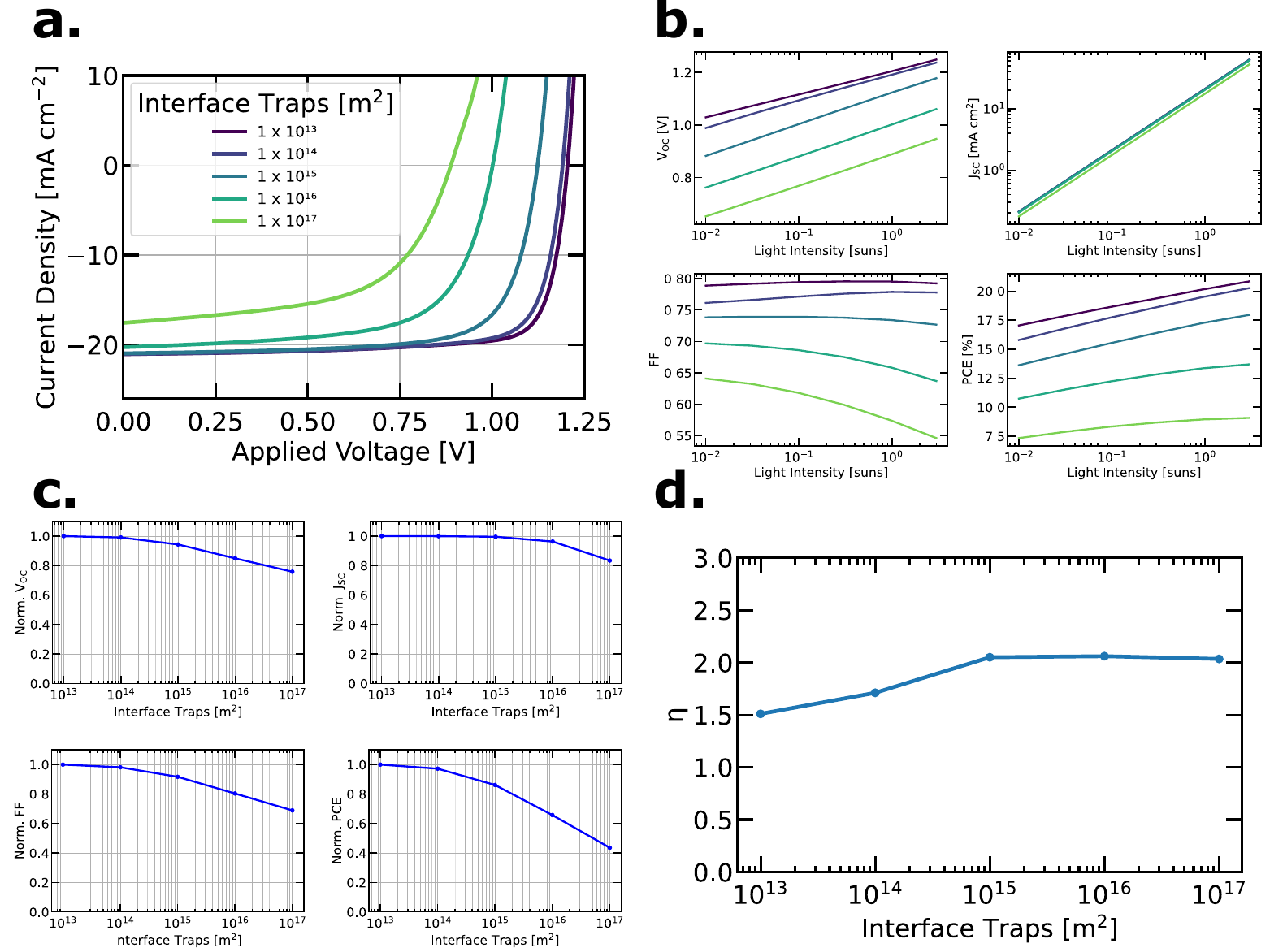}
        \caption{Influence of the interface trap density on the (a) light JVs, (b) light-intensity dependent and (c) 1 sun performances, (d) ideality factor.}\label{fig:int_trap}
\end{figure*}\FloatBarrier
\clearpage

\subsection*{OSC:}
\begin{figure*}[!htb]
    \centering
    \includegraphics[width=\textwidth]{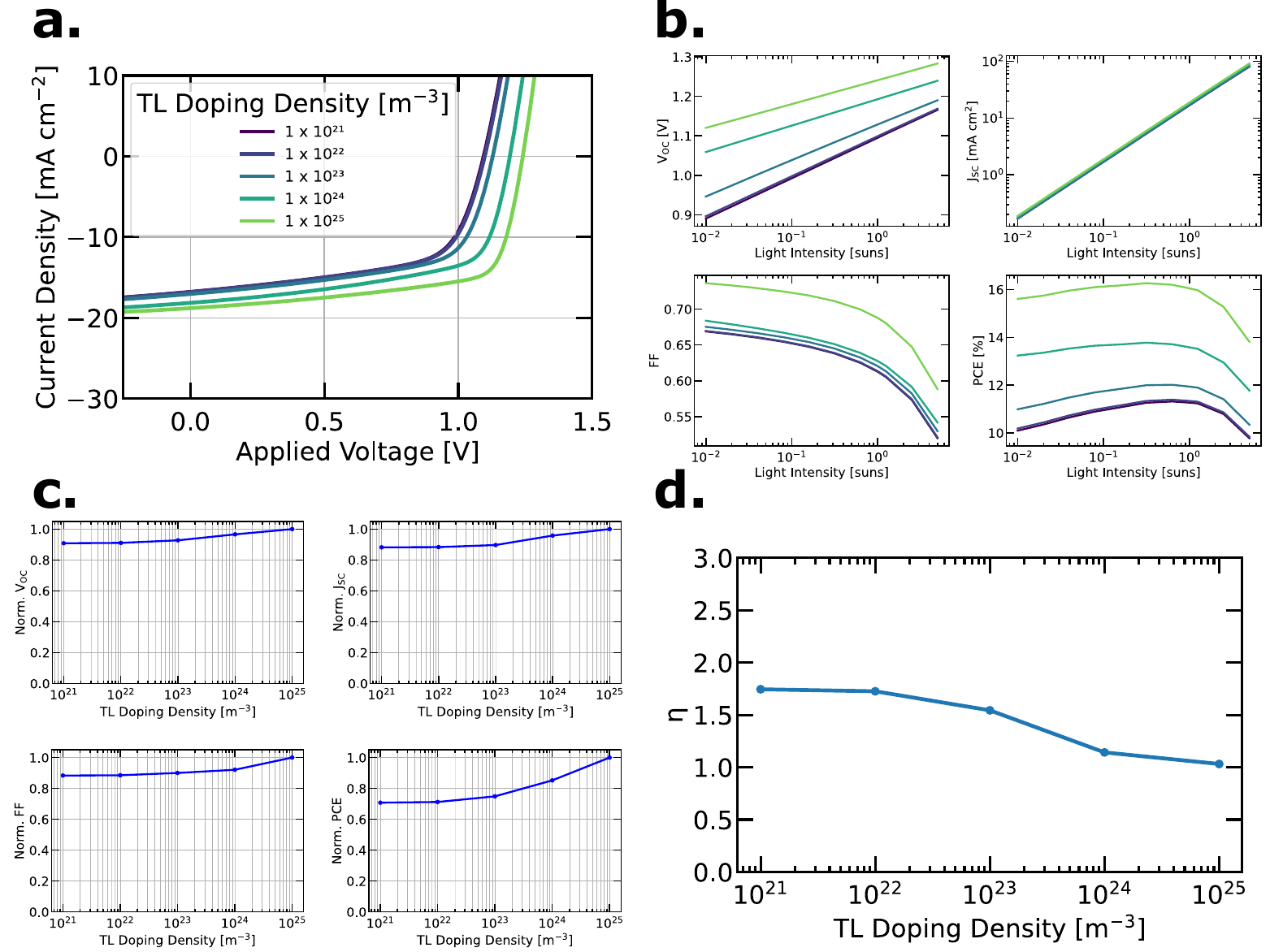}
        \caption{Influence of Fermi level pinning due to doping in the TL on the (a) light JVs, (b) light-intensity dependent and (c) 1 sun performances, (d) ideality factor.}\label{fig:OSC_pcd_dop_tl}
\end{figure*}
\begin{figure*}[!htb]
    \centering
    \includegraphics[width=\textwidth]{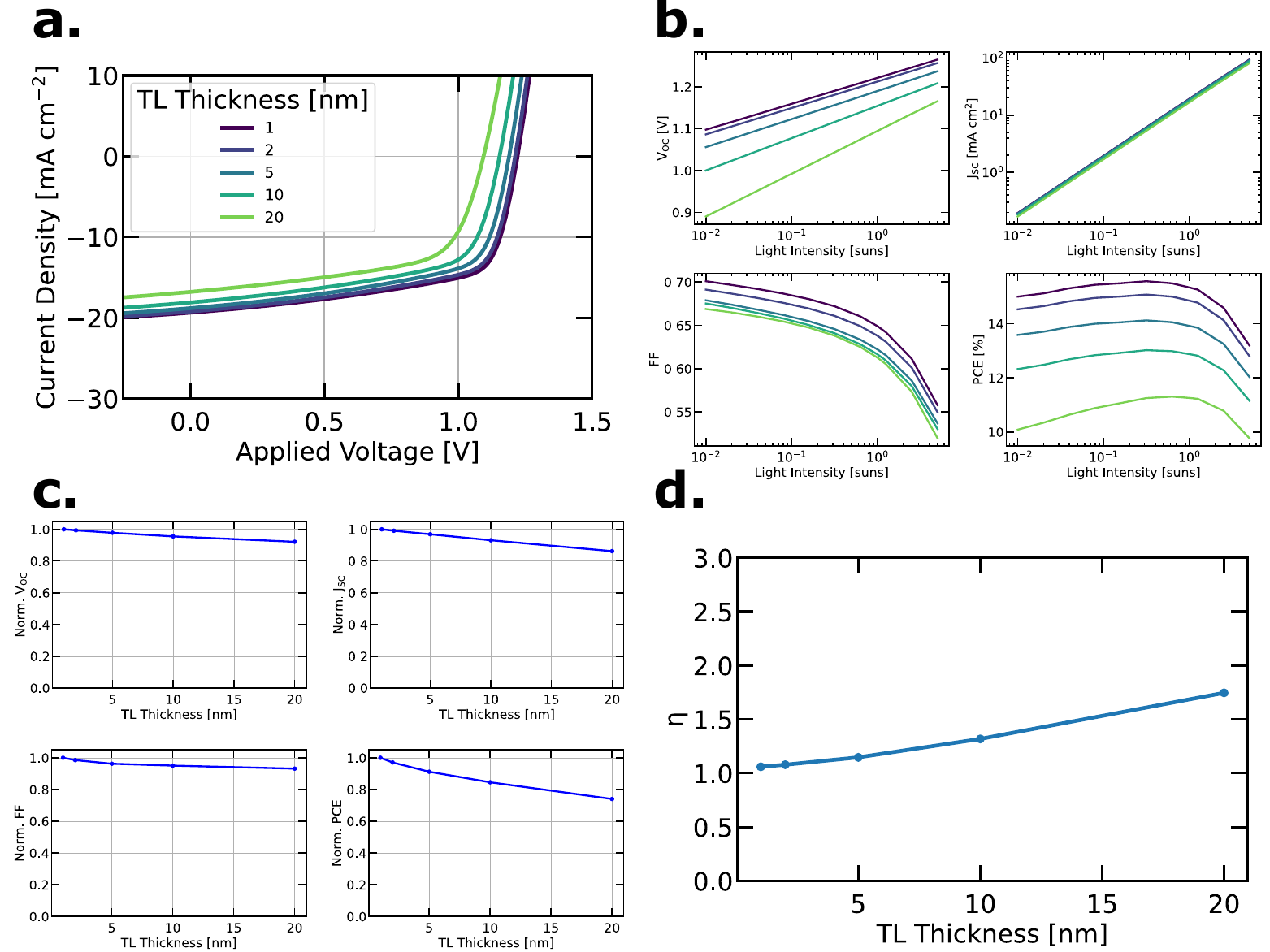}
        \caption{Influence of Fermi level pinning due TL thickness on the (a) light JVs, (b) light-intensity dependent and (c) 1 sun performances, (d) ideality factor.}\label{fig:OSC_pcd_l_tl}
\end{figure*}
\begin{figure*}[!htb]
    \centering
    \includegraphics[width=\textwidth]{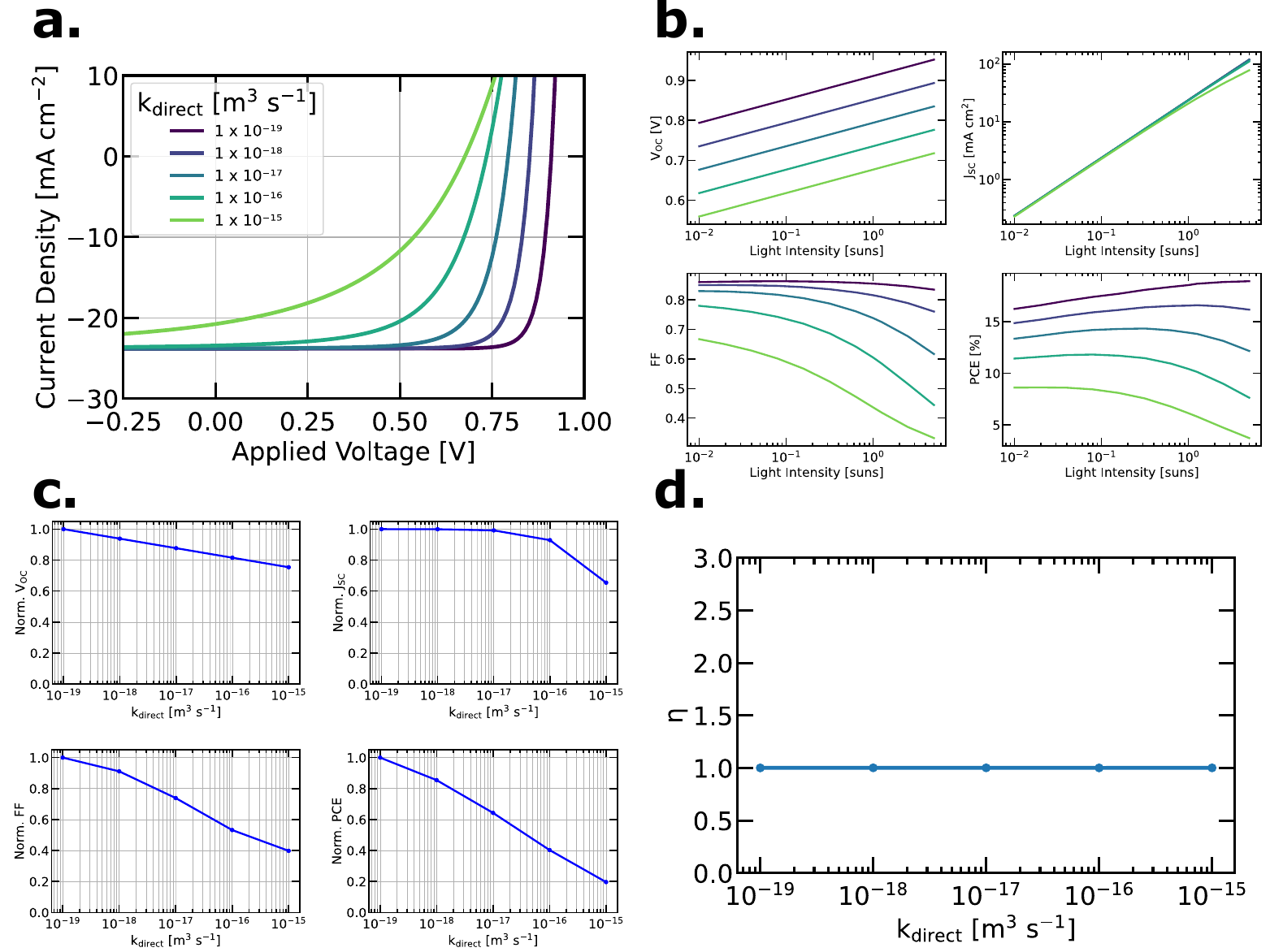}
        \caption{Influence of the band-to-band/bimolecular recombination rate ($k_2$) on the (a) light JVs, (b) light-intensity dependent and (c) 1 sun performances, (d) ideality factor.}\label{fig:OSC_k2}
\end{figure*}
\begin{figure*}[!htb]
    \centering
    \includegraphics[width=\textwidth]{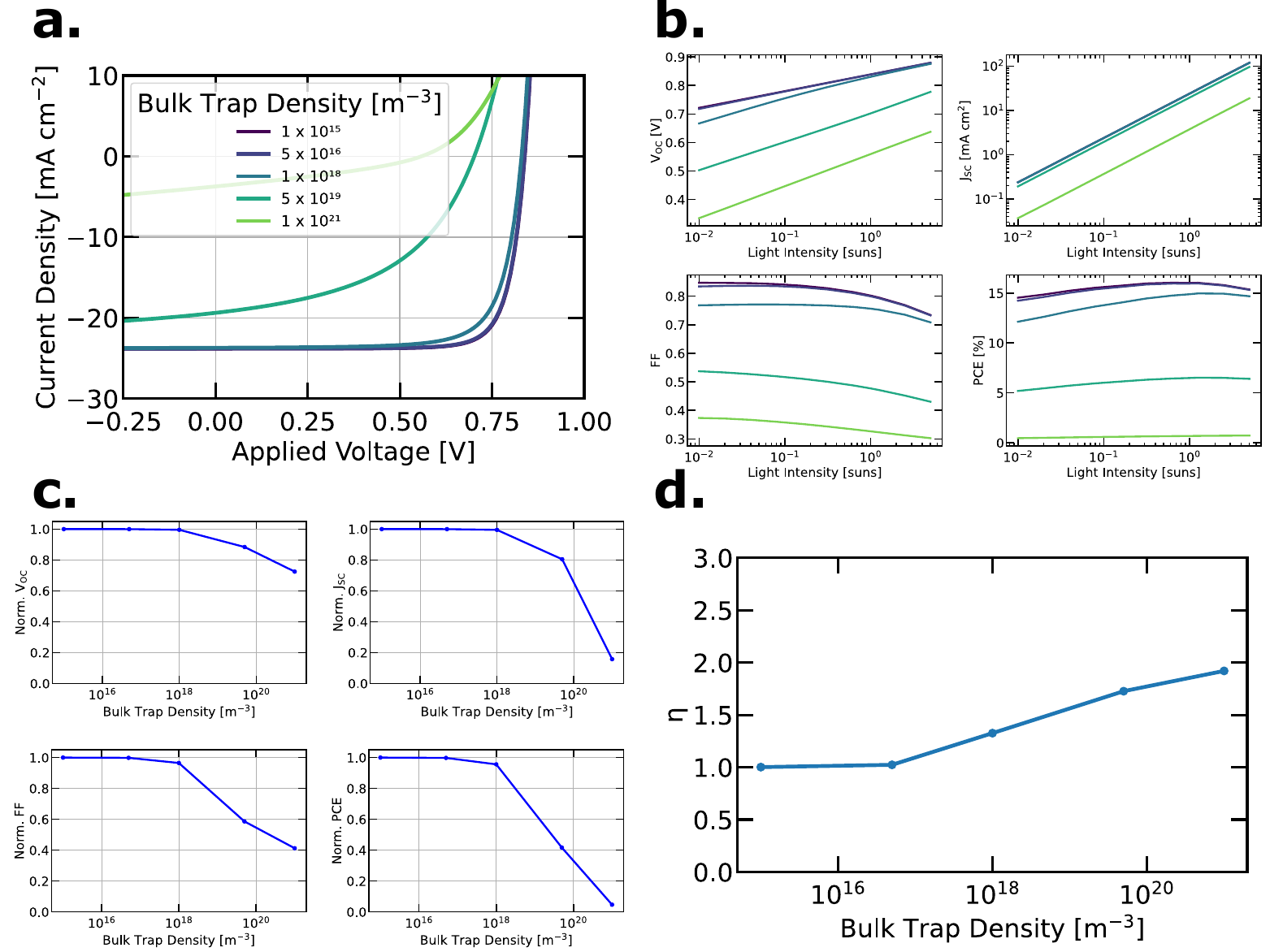}
        \caption{Influence of the bulk trap density on the (a) light JVs, (b) light-intensity dependent and (c) 1 sun performances, (d) ideality factor.}\label{fig:OSC_trap}
\end{figure*}
\begin{figure*}[!htb]
    \centering
    \includegraphics[width=\textwidth]{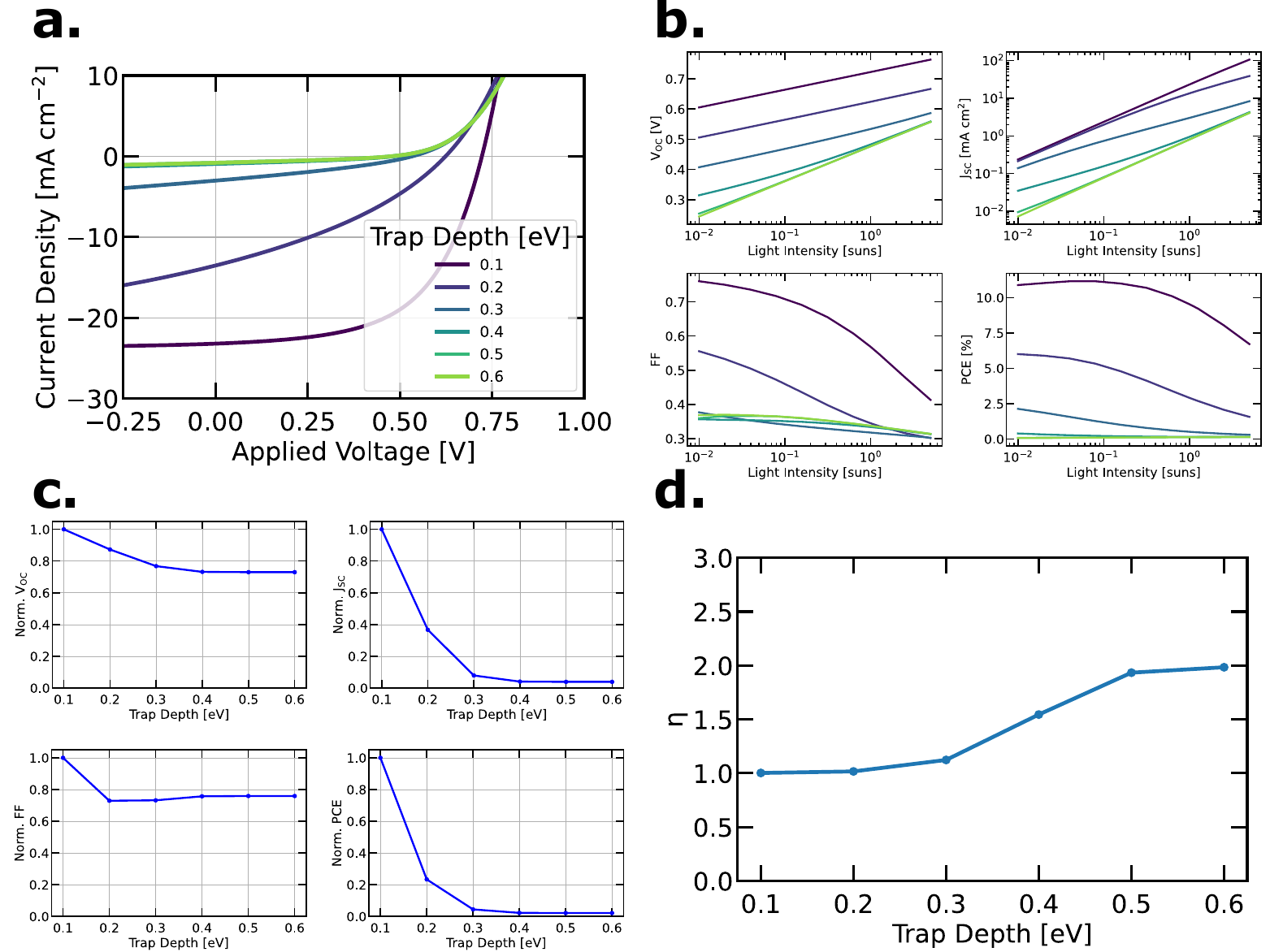}
        \caption{Influence of the trap level depth on the (a) light JVs, (b) light-intensity dependent and (c) 1 sun performances, (d) ideality factor.}\label{fig:OSC_trap_depth}
\end{figure*}

\begin{figure*}[!htb]
    \centering
    \includegraphics[width=\textwidth]{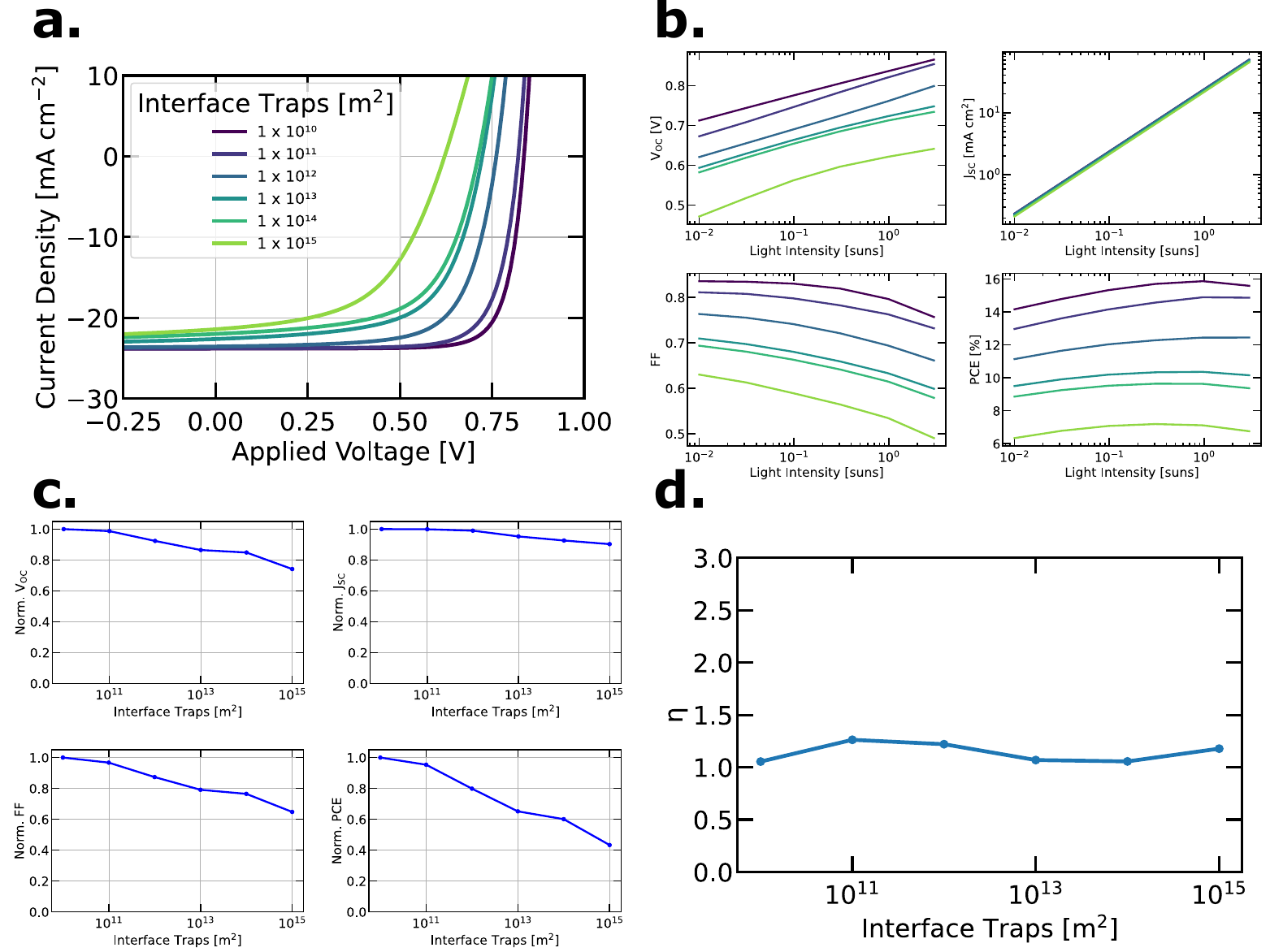}
        \caption{Influence of the interface trap density on the (a) light JVs, (b) light-intensity dependent and (c) 1 sun performances, (d) ideality factor.}\label{fig:OSC_int_trap}
\end{figure*}\FloatBarrier
\clearpage

\section{Injection barrier and trapping at the transport layer to active layer interface losses:}

\begin{figure*}[!htb]
    \centering
    \includegraphics[height=0.35\textheight]{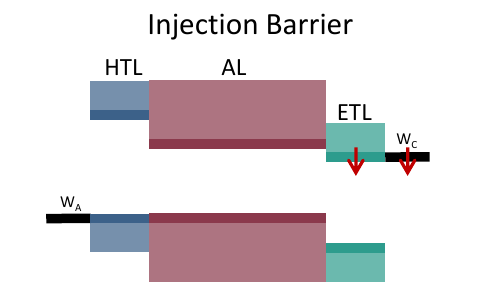}
        \caption{Sketch showing the energy band diagrams for the injection barrier}\label{fig:en_dia}
\end{figure*}\FloatBarrier
\clearpage
\subsection*{PSC:}
\begin{figure*}[!htb]
    \centering
    \includegraphics[height=0.65\textheight]{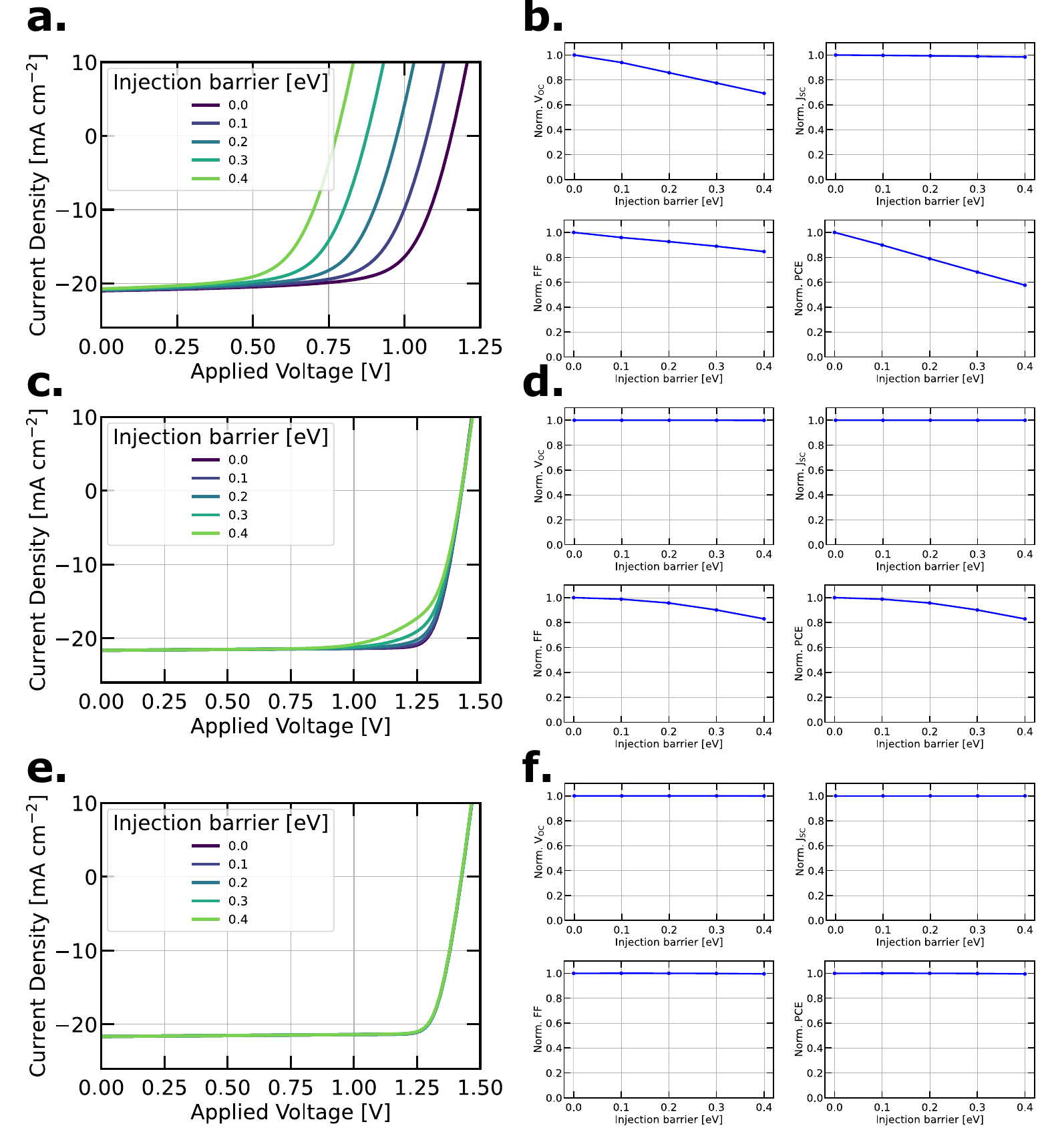}
        \caption{Influence of the injection barrier from the transport layer to the active layer with (a-b) and without (c-f) interfacial trapping on JVS and 1 sun performances. The injection barrier in the absence of trapping only reduced the $FF$, if the AL electron mobility is low (c-d), otherwise the performance is not affected (e-f)}\label{fig:injection_barrier}
\end{figure*}\FloatBarrier
\clearpage

\subsection*{OSC:}
\begin{figure*}[!htb]
    \centering
    \includegraphics[width=\textwidth]{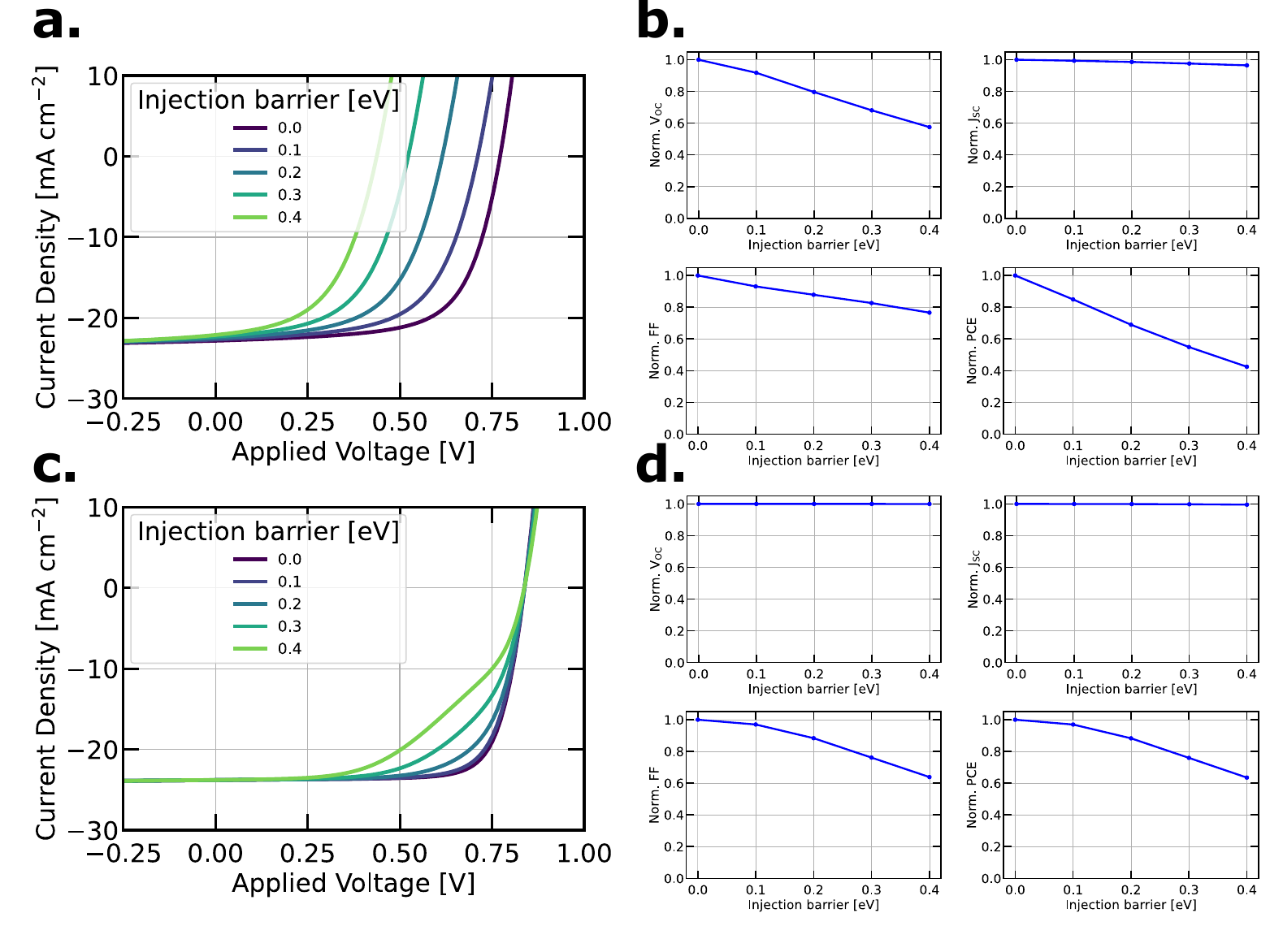}
        \caption{Influence of the injection barrier from the transport layer to the active layer with (a-b) and without (c-d) interfacial trapping on JVs and 1 sun performances. The injection barrier in the absence of trapping does reduce the $FF$ of.}\label{fig:OSC_injection_barrier}
\end{figure*}\FloatBarrier
\clearpage

\clearpage
\section{Ionic losses:}

\subsection*{PSC:}
\begin{figure*}[!htb]
    \centering
    \includegraphics[width=\textwidth]{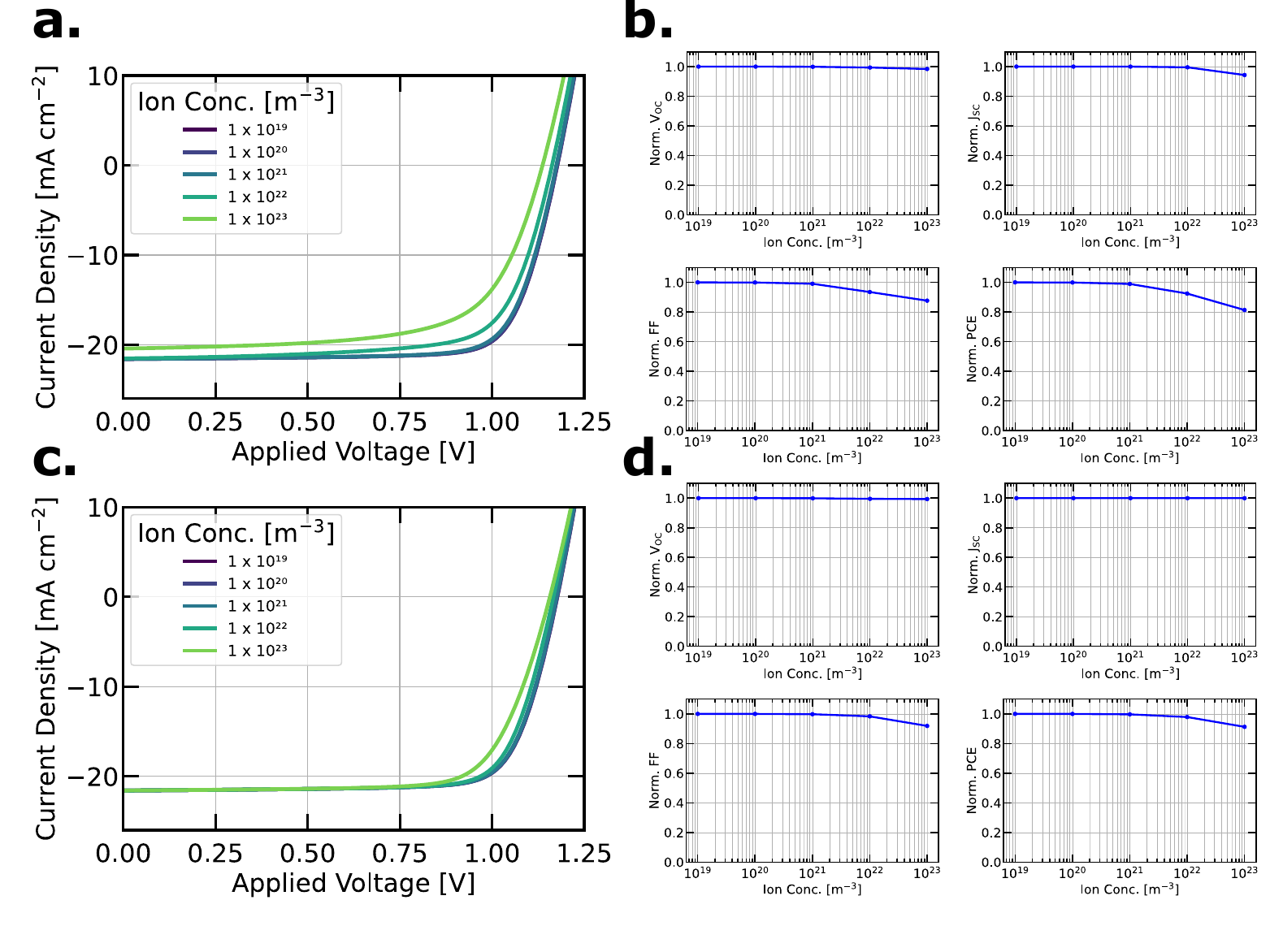}
        \caption{Influence of the concentration of positively and negatively charged ions ($I_{n} = I_{p}) $ on JVs (a,c) and performance parameters (b,d). (a-b) Ions cannot move into the transport layer, (c-d) ions can move into the transport layer.} \label{fig:ions_1}
\end{figure*}\FloatBarrier

\begin{figure*}[!htb]
    \centering
    \includegraphics[width=\textwidth]{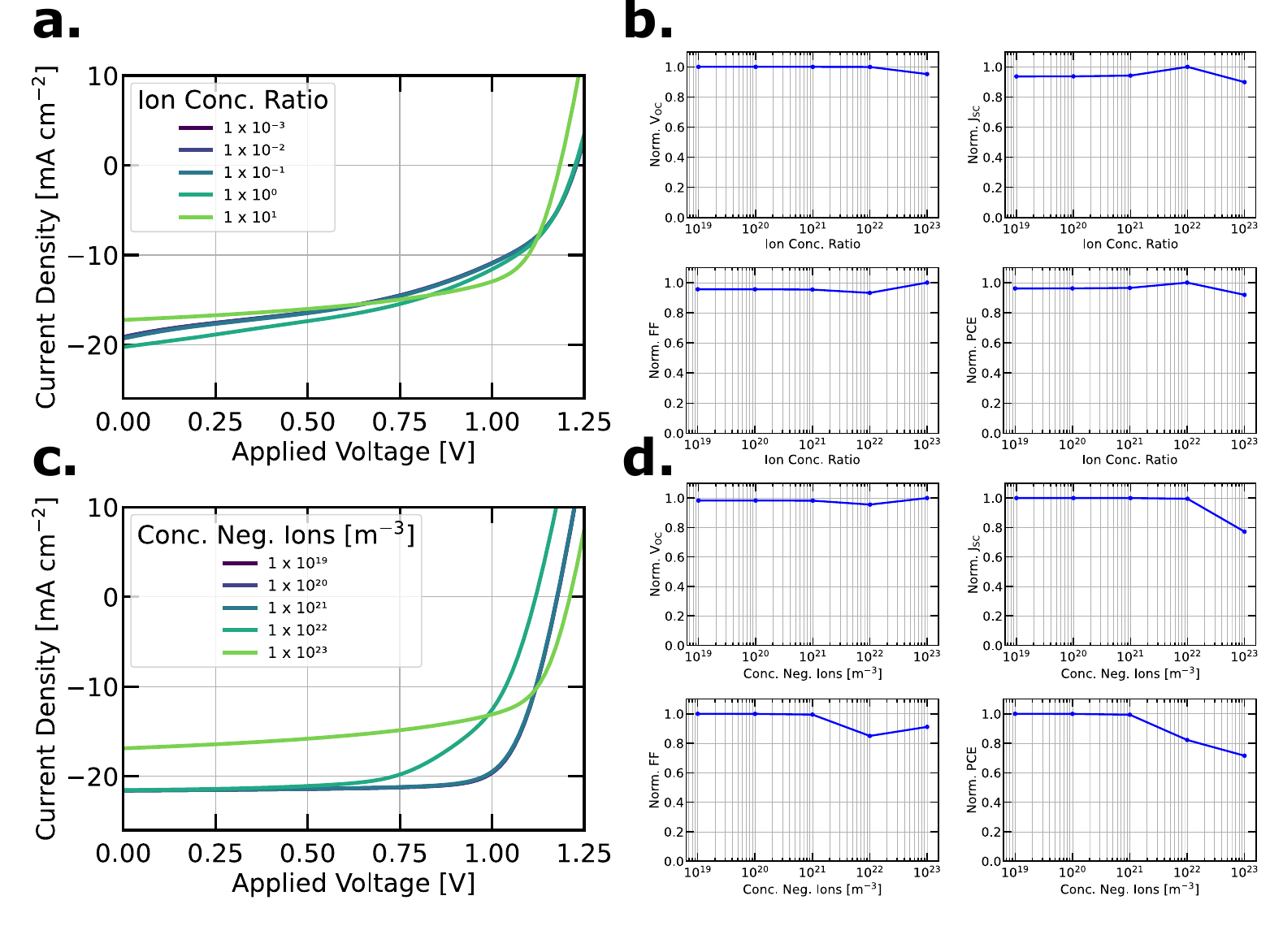}
        \caption{Influence of concentration of positively and negatively charged ions (a-b) ion ratio $\frac{I_{n}}{I_{p}}$, with $I_{p} = 1\times10^{22}$ m\textsuperscript{-3} and (c-d) $I_{n}$ with $I_{p} = 0 $ m\textsuperscript{-3} on the JVs. Ions cannot move into the transport layers.} \label{fig:ions_2}
\end{figure*}\FloatBarrier

\clearpage

\section{Field-dependant generation losses:}
\subsection*{OSC:}
\begin{figure*}[!htb]
    \centering
    \includegraphics[width=\textwidth]{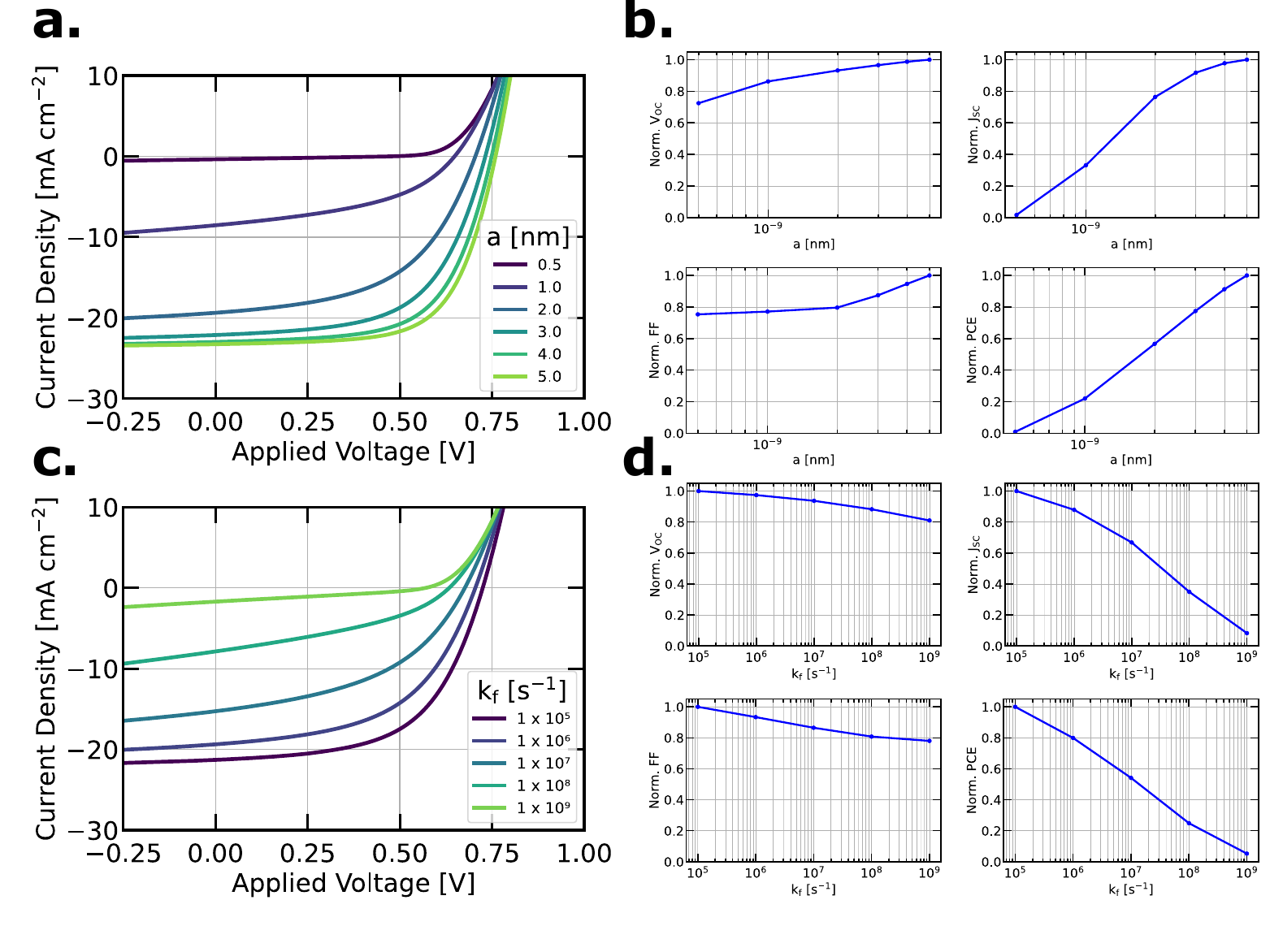}
        \caption{Influence of field-dependent charge generation losses on the JVs and 1 sun performances. (a-b)  shows the impact of the initial separation of bound charge carriers at the donor-acceptor interface $a$ and (c-d) the geminate recombination rate $k_f$.}\label{fig:OSC_fdcg}
\end{figure*}\FloatBarrier
\clearpage

\section{Series resistance losses:}
\subsection*{PSC:}
\begin{figure*}[!htb]
    \centering
    \includegraphics[width=\textwidth]{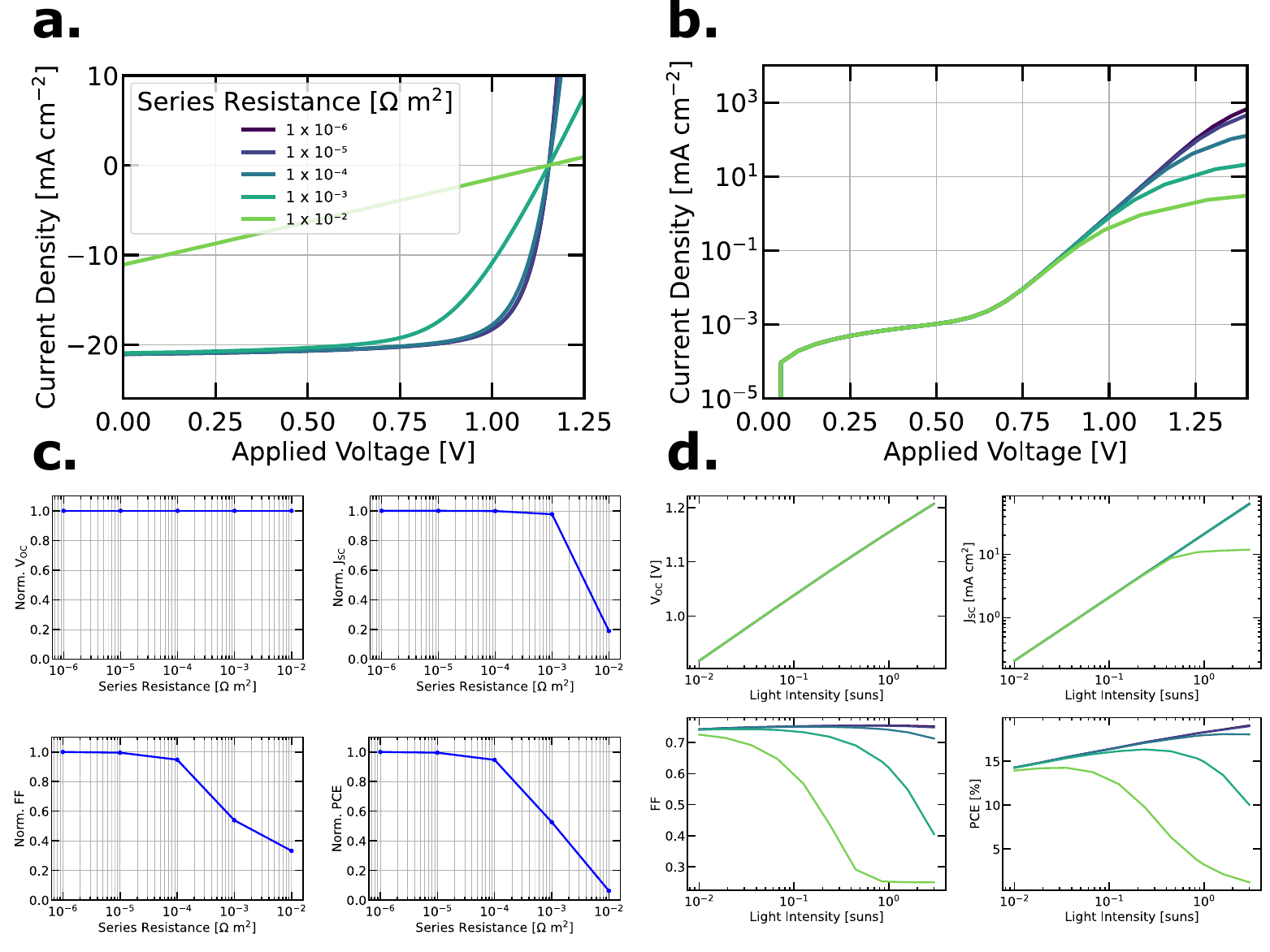}
        \caption{Influence of series resistance on the (a) light and (b) dark JVs, (c) 1 sun and (d) light-intensity dependent performances.}\label{fig:series}
\end{figure*}\FloatBarrier
\clearpage

\subsection*{OSC:}
\begin{figure*}[!htb]
    \centering
    \includegraphics[width=\textwidth]{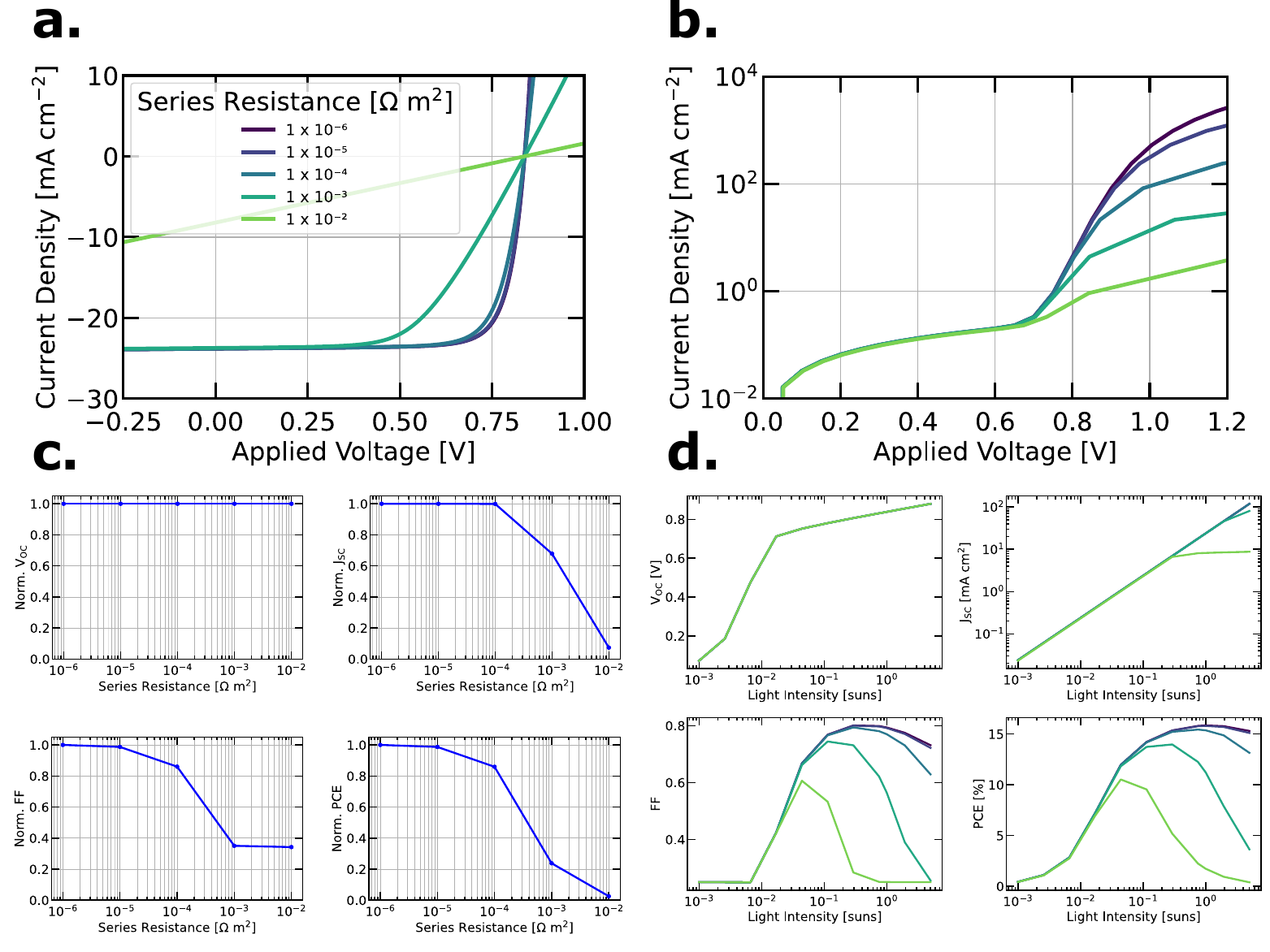}
        \caption{Influence of series resistance on the (a) light and (b) dark JVs, (c) 1 sun and (d) light-intensity dependent performances.}\label{fig:OSC_series}
\end{figure*}\FloatBarrier
\clearpage

\section{Extraction and transport losses:}
\subsection*{PSC:}
\begin{figure*}[!htb]
    \centering
    \includegraphics[width=\textwidth]{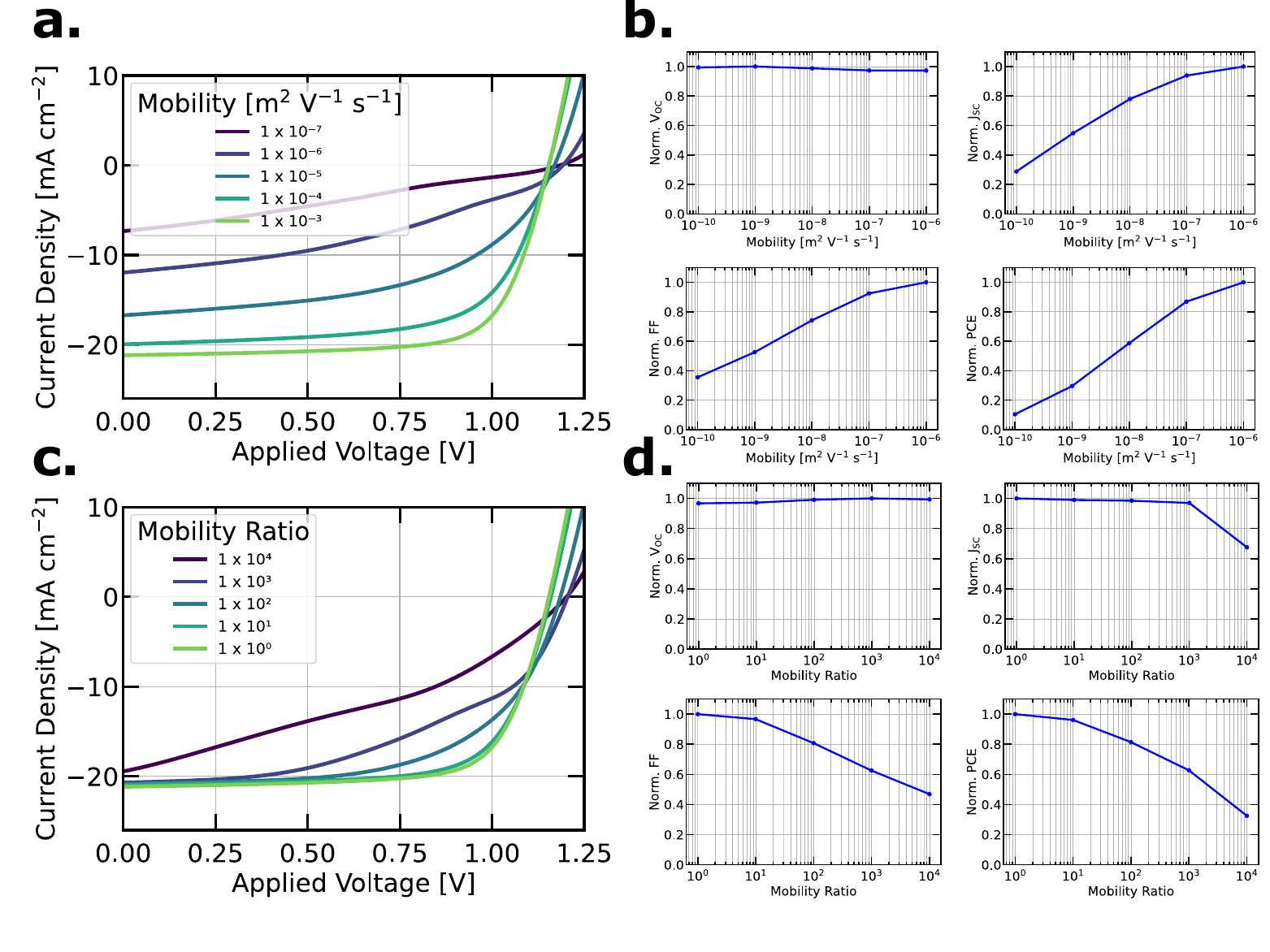}
        \caption{Influence of (a-b) active layer mobility ($\mu_n = \mu_p$) and (c-d) mobility ratio ($ratio =\frac{\mu_n}{\mu_p}$) on JVs and 1 sun performances.}\label{fig:mob}
\end{figure*}\FloatBarrier
\begin{figure*}[!htb]
    \centering
    \includegraphics[height=0.70\textheight]{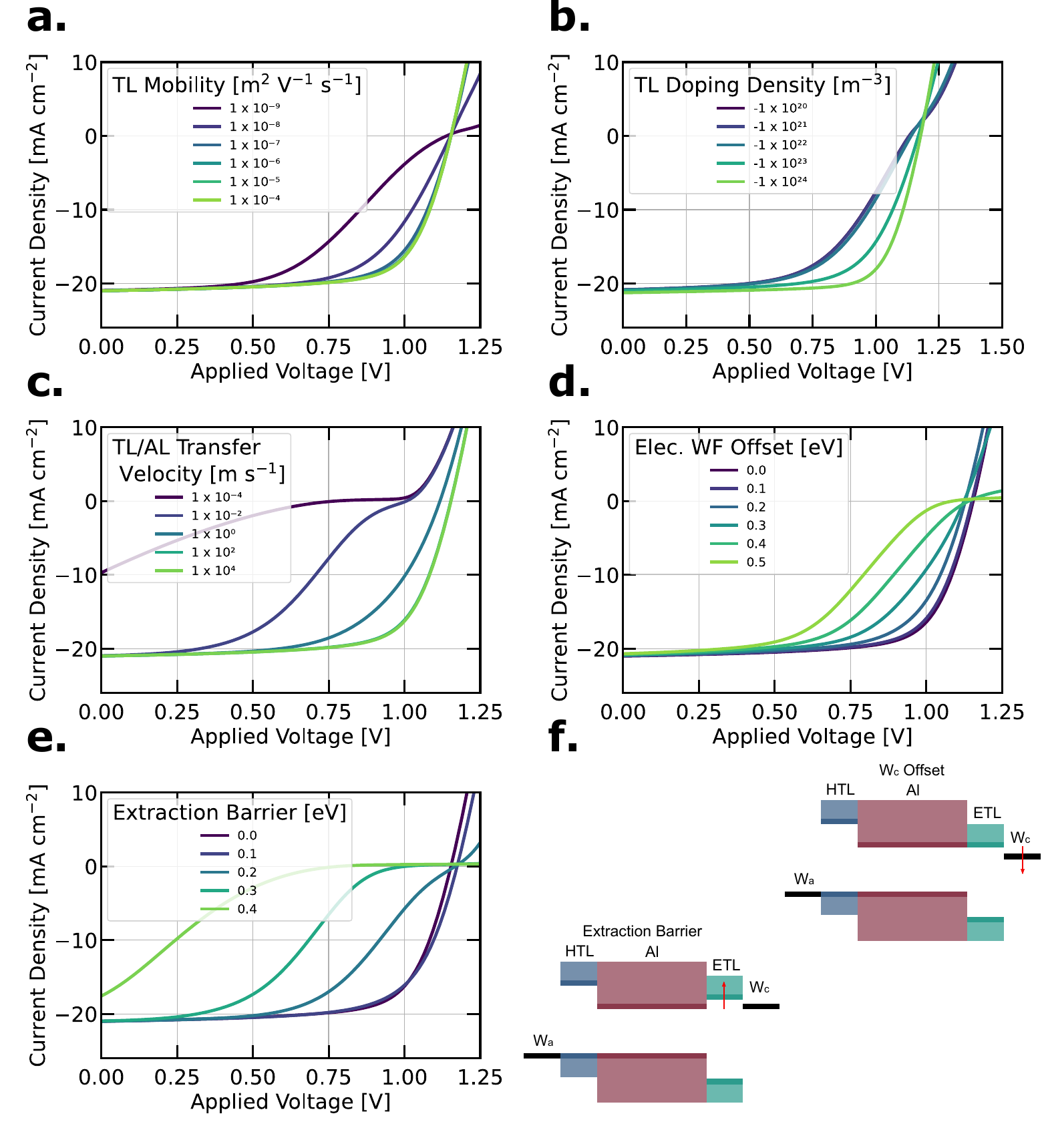}
        \caption{Influence of transport layer (a) mobility and (b) doping, (c) transfer velocity at the transport layer/active layer interface, (d) electrode work function offset (i.e. injection barrier), (e) extraction barrier from the active layer to the transport layer on the current-voltage characteristics. (f) Sketch showing the energy band diagrams showing for (d) and (e) respectively}\label{fig:s_shape}
\end{figure*}
\clearpage

\subsection*{OSC:}
\begin{figure*}[!htb]
    \centering
    \includegraphics[width=\textwidth]{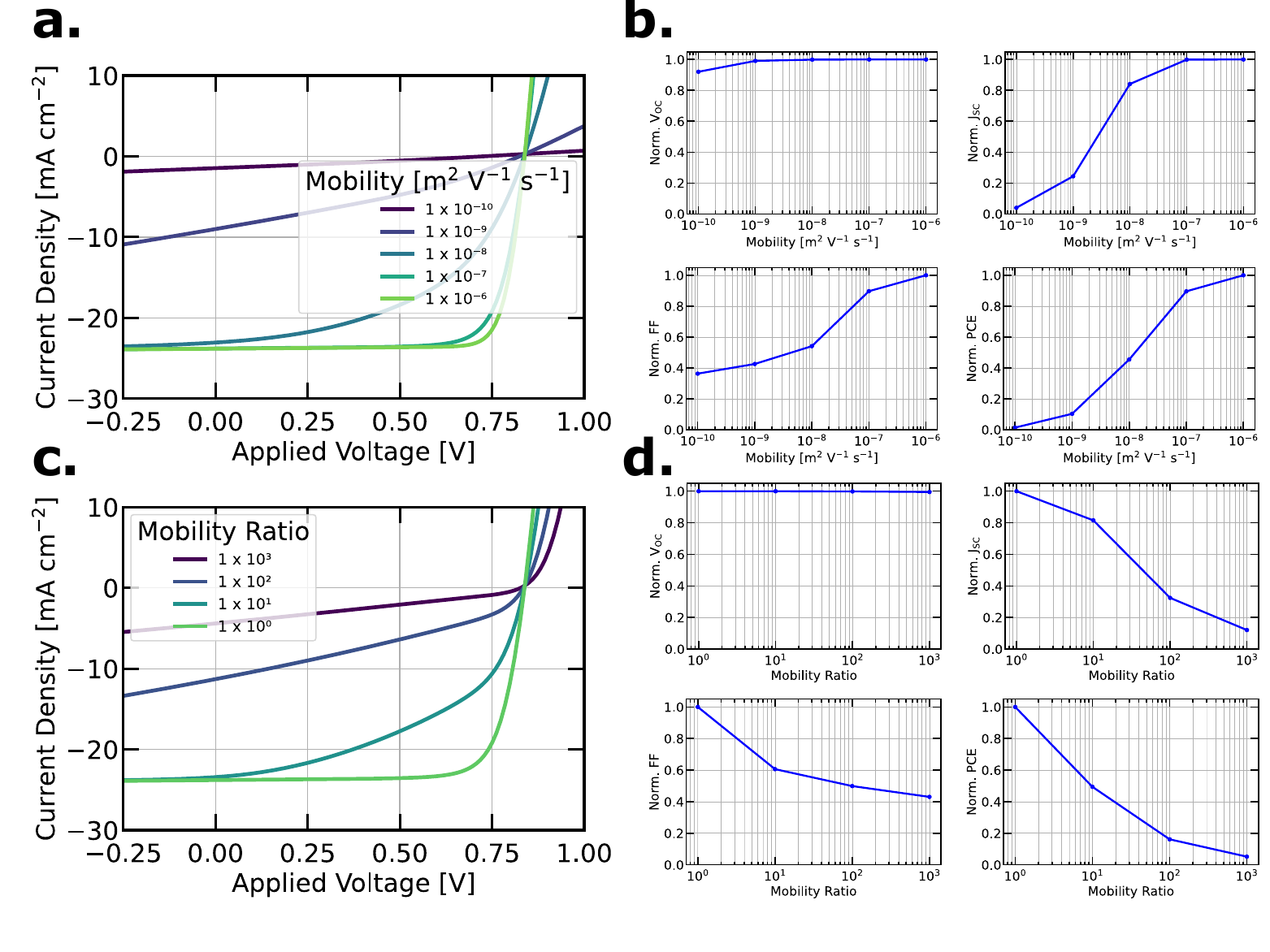}
        \caption{Influence of (a-b) active layer mobility ($\mu_n = \mu_p$) and (c-d) mobility ratio ($ratio =\frac{\mu_n}{\mu_p}$) on JVs and 1 sun performances.}\label{fig:OSC_mob}
\end{figure*}\FloatBarrier
\begin{figure*}[!htb]
    \centering
    \includegraphics[height=0.75\textheight]{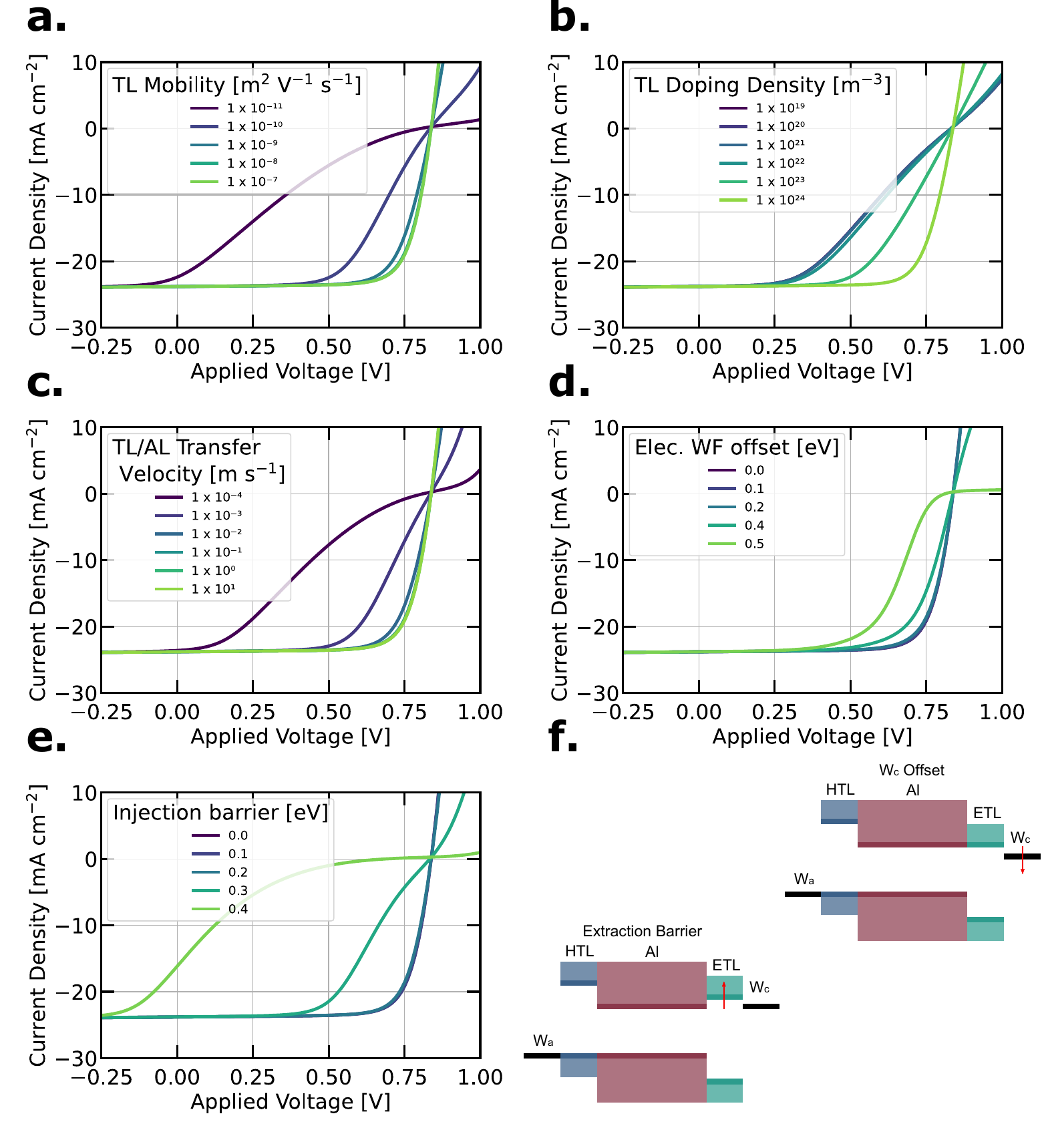}
        \caption{Influence of transport layer (a) mobility and (b) doping, (c) transfer velocity at the transport layer/active layer interface, (d) electrode work function offset (i.e. injection barrier), (e) extraction barrier from the active layer to the transport layer on the current-voltage characteristics. (f) Sketch showing the energy band diagrams showing for (d) and (e) respectively}\label{fig:OSC_s_shape}
\end{figure*}
\clearpage

\section*{Simulation Parameters:}
\begin{center}
%    Table \ref{tab:par} shows the parameters used in the base case scenario for the simulations of PSCs. 
\begin{longtable}[c]{lcccc}
	\caption{Parameters used in the base case scenario for the PSC.}\label{tab:par}\\
	% \begin{tabular}{lcccc}
		\toprule
		Parameter & Symbol & Value\\
		\midrule
		Active Layer\\
		\midrule
        Conduction band & $E_{c}$ & 3.9 eV \\
        Valence band & $E_{v}$ & 5.53 eV \\
		Band gap & $E_{gap}$ & 1.63 eV \\
		Effective density of states &  $N_{cv}$ & $1\times10^{24}$ m\textsuperscript{-3} \\
		Thickness & L & 400 nm\\
		Electron mobility & $\mu_n$ & $7\times10^{-4}$ m\textsuperscript{-2} V\textsuperscript{-1} s\textsuperscript{-1}\\
        Hole mobility & $\mu_n$ & $7\times10^{-4}$ m\textsuperscript{-2} V\textsuperscript{-1} s\textsuperscript{-1}\\
		Relative dielectric constant & $\epsilon_r$ & 22\\
        Concentration of negative (positive) ions & $I_{n(p)}$ & $6\times10^{22}$ m\textsuperscript{-3} \\
		Initial charge separation distance & $a$ &$1\times10^{-9}$ m \\
        \midrule
		Electron transport layer \\
		\midrule
		Thickness & $L^{ETL}$ & 30 nm \\
		Mobility & $\mu^{ETL}$ & $1\times10^{-6}$ m\textsuperscript{-2} V\textsuperscript{-1} s\textsuperscript{-1}\\
		Relative dielectric constant & $\epsilon_r^{ETL}$ & 5\\
		Conduction band & $E_{c}^{ETL}$ & 3.9 eV \\
        Valence band & $E_{v}^{ETL}$ & 5.9 eV \\
        Doping density & $N^+_D$ & $0$ m\textsuperscript{-3}\\
        \midrule
		Hole transport layer \\
		\midrule
		Thickness & $L^{HTL}$ & 10 nm \\
		Mobility & $\mu^{HTL}$ & $1.5\times10^{-8}$ m\textsuperscript{-2} V\textsuperscript{-1} s\textsuperscript{-1}\\
		Relative dielectric constant & $\epsilon_r^{HTL}$ & 3.5\\
		Conduction band & $E_{c}^{HTL}$ & 2.5 eV \\
        Valence band & $E_{v}^{HTL}$ & 5.53 eV \\
        Doping density & $N^-_A$ & $0$ m\textsuperscript{-3}\\
		\midrule
		Generation \& Recombination\\
		\midrule
        Average generation rate & $Gehp$ & $3.4\times10^{27}$ m\textsuperscript{-3} s\textsuperscript{-1}\\
		Band-to-band/Bimolecular recombination rate& $k_2$ & $1\times10^{-17}$ m\textsuperscript{3} s\textsuperscript{-1}\\
        Bulk trap density & $N_T$ & $5\times10^{21}$ m\textsuperscript{-3}\\
        ETL/AL interface trap density & $\Sigma_{T}^{ETL}$ & $5\times10^{14}$ m\textsuperscript{-2}\\
        HTL/AL interface trap density & $\Sigma_{T}^{HTL}$ & $5\times10^{13}$ m\textsuperscript{-2}\\
        Trap energy level & $E_{T}$ & 4.7 eV\\
        Electron (hole) capture coefficient & $C_{n(p)}$ & $1\times10^{-14}$ m\textsuperscript{3} s\textsuperscript{-1}\\
		Geminate recombination rate & $k_f$ & $1\times10^{6}$ s\textsuperscript{-1}\\
        \midrule
		Contact\\
		\midrule
		Cathode work function & $W_L$ & 3.95 eV\\
        Anode work function & $W_R$ & 5.48 eV\\
		\bottomrule	
	% \end{tabular}
\end{longtable}

% Table \ref{tab:par_OSC} shows the parameters used in the base case scenario for the simulations of OSCs. 
\begin{longtable}[c]{lcccc}
	\caption{Parameters used in the base case scenario for the OSC.}\label{tab:par_OSC}\\
	% \begin{tabular}{lcccc}
		\toprule
		Parameter & Symbol & Value\\
		\midrule
		Active Layer\\
		\midrule
        Conduction band & $E_{c}$ & 4.2 eV \\
        Valence band & $E_{v}$ & 5.42 eV \\
		Band gap & $E_{gap}$ & 1.22 eV \\
		Effective density of states &  $N_{cv}$ & $1.55\times10^{26}$ m\textsuperscript{-3} \\
		Thickness & L & 120 nm\\
		Electron mobility & $\mu_n$ & $9.32\times10^{-8}$ m\textsuperscript{-2} V\textsuperscript{-1} s\textsuperscript{-1}\\
        Hole mobility & $\mu_n$ & $9.78\times10^{-8}$ m\textsuperscript{-2} V\textsuperscript{-1} s\textsuperscript{-1}\\
		Relative dielectric constant & $\epsilon_r$ & 3.5\\
		Initial charge separation distance & $a$ &$1\times10^{-9}$ m \\
        \midrule
		Electron transport layer \\
		\midrule
		Thickness & $L^{ETL}$ & 10 nm \\
		Mobility & $\mu^{ETL}$ & $1\times10^{-6}$ m\textsuperscript{-2} V\textsuperscript{-1} s\textsuperscript{-1}\\
		Relative dielectric constant & $\epsilon_r^{ETL}$ & 3.5\\
		Conduction band & $E_{c}^{ETL}$ & 4.2 eV \\
        Valence band & $E_{v}^{ETL}$ & 6. eV \\
        Doping density & $N^+_D$ & $0$ m\textsuperscript{-3}\\
        \midrule
		Hole transport layer \\
		\midrule
		Thickness & $L^{HTL}$ & 40 nm \\
		Mobility & $\mu^{HTL}$ & $1\times10^{-6}$ m\textsuperscript{-2} V\textsuperscript{-1} s\textsuperscript{-1}\\
		Relative dielectric constant & $\epsilon_r^{HTL}$ & 3.5\\
		Conduction band & $E_{c}^{HTL}$ & 3 eV \\
        Valence band & $E_{v}^{HTL}$ & 5.42 eV \\
        Doping density & $N^-_A$ & $0$ m\textsuperscript{-3}\\
		\midrule
		Generation \& Recombination\\
		\midrule
        Average generation rate & $Gehp$ & $1.3\times10^{28}$ m\textsuperscript{-3} s\textsuperscript{-1}\\
		Band-to-band/Bimolecular recombination rate& $k_2$ & $1.7\times10^{-18}$ m\textsuperscript{3} s\textsuperscript{-1}\\
        Bulk trap density & $N_T$ & ${0}$ m\textsuperscript{-3}\\
        ETL/AL interface trap density & $\Sigma_{T}^{ETL}$ & $0$ m\textsuperscript{-2}\\
        HTL/AL interface trap density & $\Sigma_{T}^{HTL}$ & $0$ m\textsuperscript{-2}\\
        Trap energy level & $E_{T}$ & 4.7 eV\\
        Electron (hole) capture coefficient & $C_{n(p)}$ & $1\times10^{-13}$ m\textsuperscript{3} s\textsuperscript{-1}\\
		Geminate recombination rate & $k_f$ & $1\times10^{6}$ s\textsuperscript{-1}\\
        \midrule
		Contact\\
		\midrule
		Cathode work function & $W_L$ & 4.2 eV\\
        Anode work function & $W_R$ & 5.42 eV\\
		\bottomrule	
	% \end{tabular}
\end{longtable}
\end{center}

\section*{Influence of parasitic resistances on fill factor and open-circuit voltage}

In the following, we derive two criteria to quickly assess whether measurements of the fill factor and open-circuit voltage are influenced by shunt and series resistance. This is especially important when doing such measurements as a function of light intensity.

\subsection*{Shunt resistance}

Leakage current will influence the fill factor and open-circuit voltage. This is especially important at low light intensities. Therefore, one needs a solid way of determining whether measurements of $FF$ or $V_{OC}$ are significantly influenced by leakage. In the following, we accept a 1 \% deviation.\\
We can set up such a criterion for $FF$ and $V_{OC}$ by considering an equivalent circuit. According to Martin Green’s estimate,\cite{Green1981} the fill factor in the presence of leakage (finite shunt resistance $R_{SH}$) is given by 

\begin{equation}\label{FF_Green}
    FF = FF_0(1-\frac{v_{OC}+0.7}{v_{OC}}\frac{FF_0}{r_{SH}}),
\end{equation}
where $FF_0$ is the fill factor in absence of leakage (the ’real’ fill factor), $v_{OC}=\frac{V_{OC}}{nV_{th}}$ is the scaled open-circuit voltage, and $r_{SH}=\frac{R_{SH} J_{SC}}{V_{OC}}$. Typically, $V_{OC}$ $\gg$  $V_{th}$, so the term containing $V_{OC}$ is approximately 1. Also, for good cells, $FF$ is close to 1, so a 1\% deviation equals 0.01.\footnote{Green’s formulae are limited to good solar cells anyway.} This means that $FF$ will be unaffected provided $\frac{FF_0}{R_{SH}} < 0.01$. The shunt resistance may be approximated by

\begin{equation}\label{Rsh_approx}
    R_{SH} \approx \frac{1 V}{J_{dark}(-1V)}.
\end{equation}
Finally, we find that $FF$ is not significantly affected (less than 1\%) if

\begin{equation}\label{shunt_ff}
    \frac{J_{SC}}{J_{dark}(-1V)} \gtrsim 100.
\end{equation}

\begin{figure*}[!htb]
    \centering
    \includegraphics[width=\textwidth]{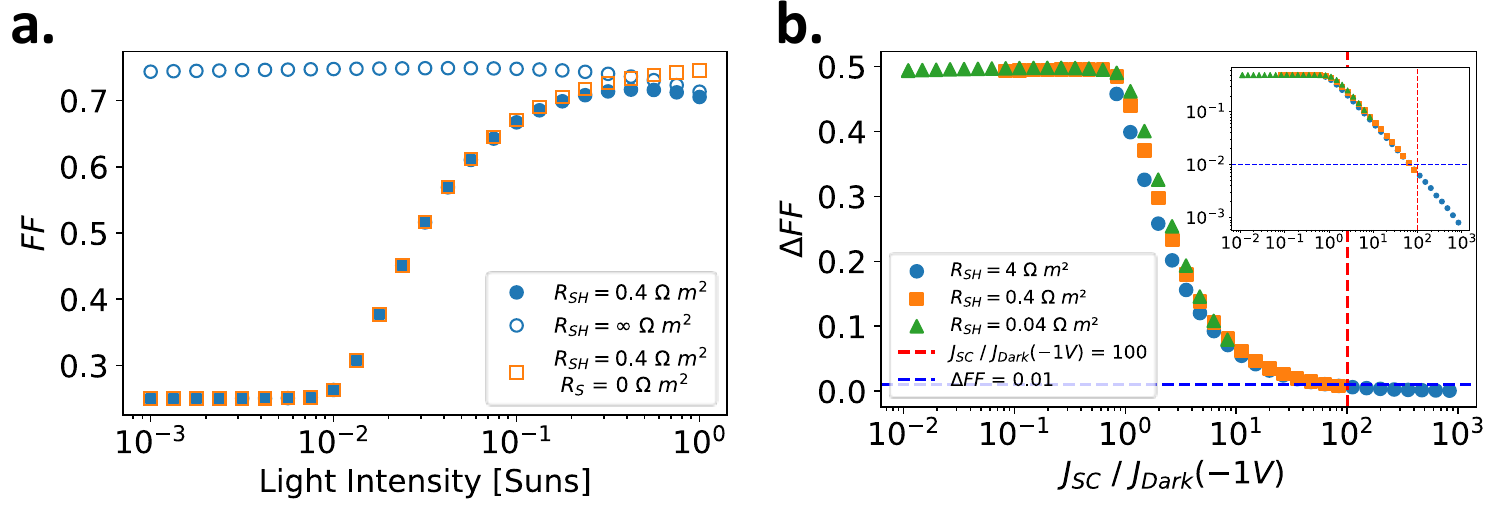}
        \caption{(a) The fill factor of the perovskite base scenario with finite shunt resistance (closed circles) and with infinite shunt resistance (open circles). The effect of removing series resistance is shown by including simulations with zero series resistance (orange rectangles). (b) Shows the deviation of the fill factor as a function of the ratio between the short-circuit current and the leakage current for three different shunt resistances. The dashed red vertical line indicates a ratio of 100 (where the criterion starts to apply) and the blue horizontal line a deviation of the $FF$ of 1 \%. The fill factor is thus impacted by leakage unless the ratio is larger than $\approx$ 100. The inset depicts the same plot in a log-log scale, where one can  see that the fill factor deviates by more than 0.01 if the ratio is smaller than 100.}\label{fig:FF_shunt}
\end{figure*}

To illustrate the effect of shunt resistance on $FF$ and how to use the criterion Eq.~\ref{shunt_ff}, we use the perovskite reference device (Table \ref{tab:par}) and simulate the fill factor as a function of light intensity. Figure \ref{fig:FF_shunt} demonstrates that Eq.~\ref{shunt_ff} indeed holds. In other words, the fill factor is quite sensitive to leakage current and is impacted unless the light intensity is high enough to ensure that the short-circuit current is much larger than the leakage.

\FloatBarrier

Now we turn to the open-circuit voltage. We will find that $V_{OC}$ is less affected by shunt resistance, despite the higher bias (and therefore higher leakage current). The current in an equivalent circuit that includes shunt resistance is equal to

\begin{equation}\label{EC}
    J(V) = J_{SC} - J_0\exp(\frac{V}{nV_{th}}) - \frac{V}{R_{SH}},
\end{equation}
where we have neglected the +1 term in the exponential. At open-circuit
the current density is zero and we have

\begin{equation}\label{EC_OC}
    J_{SC} - \frac{V}{R_{SH}} = J_0\exp(\frac{V}{nV_{th}})
\end{equation}
If we use the normal result for $V_{OC}$ in the absence of parasitic resistances, and replace $J_{SC}$ by $J_{SC} - \frac{V_{OC}}{R_{SH}}$, we have

\begin{equation}\label{VOC_approx}
    V_{OC} \approx nV_{th}\ln(\frac{J_{SC}-1/R_{SH}}{J_0}). 
\end{equation}
The deviation of $V_{OC}$ due to leakage current is given by

\begin{equation}\label{VOC_influence_rsh}
    \Delta V_{OC} = nV_{th} \bigl\{ \ln \ J_{SC} - \ln \bigl( J_{SC} - \frac{1}{R_{SH}} \bigr) \bigr\}.
\end{equation}
Using a first order Taylor series yields

\begin{equation}\label{Taylor}
    \Delta V_{OC} = nV_{th}\frac{J_{dark}(-1 V)}{J_{SC}}.
\end{equation}

\begin{figure*}[!htb]
    \centering
    \includegraphics[width=\textwidth]{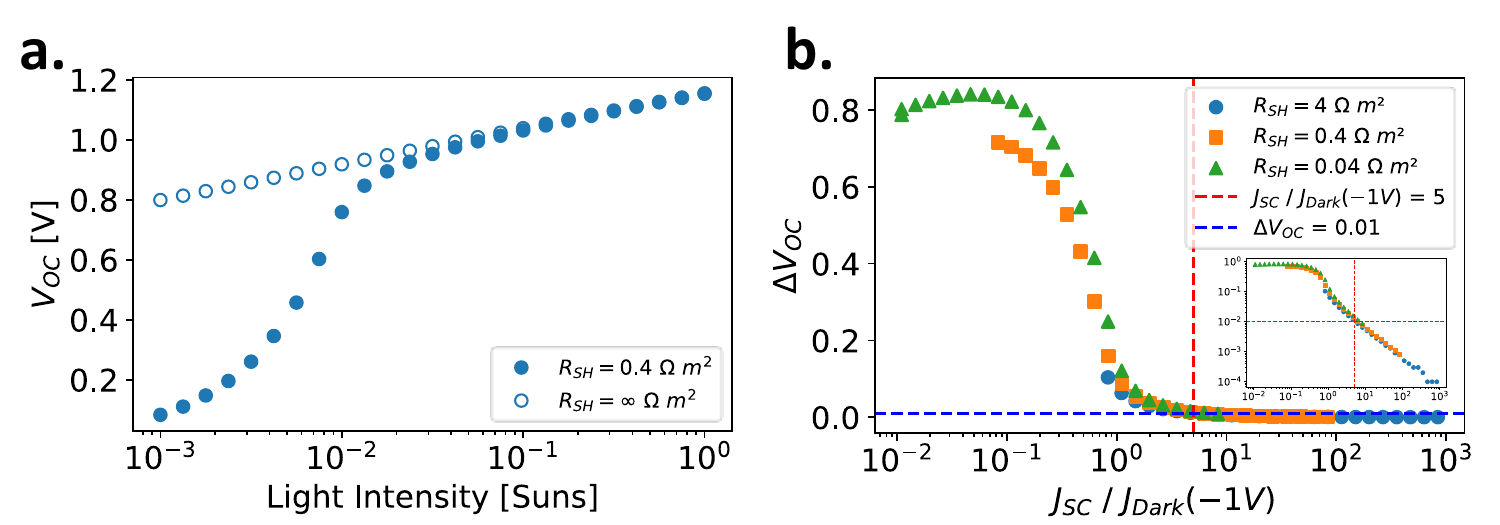}
        \caption{(a) Open-circuit voltage of the perovskite base scenario with finite shunt resistance (closed symbols) and with infinite shunt resistance (open symbols). (b) Shows the deviation of the open-circuit voltage as a function of the ratio between the short-circuit current and the leakage current for three different shunt resistances. The dashed red vertical line indicates a ratio of 5 (where the criteria starts to apply) and the blue horizontal line a deviation of the $V_{OC}$ of 0.01. The inset depicts the same plot in a log-log scale, where one can see that the open-circuit voltage deviates by more than 0.01 V if the ratio is smaller than 5 (see Eq.~\ref{shunt_voc}).}\label{fig:Voc_shunt}
\end{figure*}

Now we require $\Delta V_{OC}$ be smaller than 0.01 V (i.e. 1\%) and obtain

\begin{equation}\label{shunt_voc_thermal}
    \frac{J_{SC}}{J_{dark}(-1V)} \gtrsim 100nV_{th},
\end{equation}
which is very similar to Eq. \ref{shunt_ff}, the only difference being the $nV_{th}$ term. At room temperature and for $n \approx 2 $, we have

\begin{equation}\label{shunt_voc}
    \frac{J_{SC}}{J_{dark}(-1V)} \gtrsim 5,
\end{equation}
which shows that the open-circuit voltage is indeed less sensitive to leakage. In other words, measurements of $V_{OC}$ versus light intensity will be correct down to lower intensities than similar measurements of $FF$. Figure \ref{fig:Voc_shunt} shows the influence of leakage on the open-circuit voltage. Indeed, if the criterion (Eq.~\ref{shunt_voc_thermal}) is satisfied, the deviation in open-circuit is below 0.01 V. 

\FloatBarrier

\subsection*{Series resistance}

If there is a noticeable amount of series resistance in the equivalent circuit of the device, then this will impact the fill factor. The short-circuit current density will only change, if the series resistance is very high. The open-circuit voltage is not impacted by series resistance as there is no flow of current and, hence, no change in voltage. Therefore, we limit our discussion to mild series resistance only, i.e. the case where $FF$ changes, but the short-circuit current does not. In order to derive how $FF$ changes by series resistance, we assume that the strongest effect on the maximum power point is in the shift of its voltage ($V_{MPP}$), whereas its current ($J_{MPP}$) is assumed to be unaffected. Then, we can write $FF$ as

\begin{equation}\label{ser_FF}
    FF = \frac{J_{MPP} (V_{MPP}-J_{MPP}R_s)}{(J_{SC}V_{OC})}.
\end{equation}
The change in $FF$, $\Delta FF$ is then equal to

\begin{equation}\label{ser_delta_FF}
    \Delta FF = \frac{J_{SC}R_s}{V_{OC}},
\end{equation}
where we have approximated $J_{SC}$ by $J_{MPP}$. As $J_{MPP}$ is smaller than $J_{SC}$, Eq. \ref{ser_delta_FF} overestimates the change in $FF$. If we, again, accept an error of 0.01, then we have the criterion that $FF$ is valid is
\begin{equation}
    J_{SC} < \frac{0.01 V_{OC}}{R_s}.
\end{equation}
Alternatively, if we take $V_{OC}\approx 1$V, then we have that
\begin{equation}
    J_{SC} R_s < 0.01 {\rm V}.
\end{equation}
Figure \ref{fig:FF_shunt} shows the impact of series resistance on the fill factor.

\clearpage
\bibliographystyle{vlcref}
\bibliography{guide}